\documentclass{ws-procs9x6}
\usepackage{latexsym}
\usepackage{epsfig,graphics}

\newcommand{\be}{\begin{equation}}
\newcommand{\ee}{\end{equation}}

\begin{document}

\title{Large-N gauge theories: \\
lattice perspectives and conjectures} 

\author{M. TEPER}

\address{Rudolf Peierls Centre for Theoretical Physics, \\
University of Oxford, \\ 
1 Keble Road, \\
Oxford OX1 3NP, UK\\ 
E-mail: teper@thphys.ox.ac.uk}

\maketitle

\abstracts{
I summarise what recent lattice calculations tell
us about the large-$N$ limit of SU($N$) gauge theories in
3+1 dimensions. The focus is on confinement, how close SU($\infty$) is
to SU(3), new stable strings at larger $N$, 
deconfinement, topology and $\theta$-vacua. I discuss the effective
string theory description, as well as master fields,
space-time reduction and non-analyticity.}

\section{Introduction, Overview, Conjectures}
\label{section_Introduction}

It is 30 years since it was proposed that it might 
be useful to think of QCD as a perturbation in $1/N$ around
the $N=\infty$ theory 
\cite{thooftN}. 
As is apparent from the talks at this meeting, 
this has been a very fruitful idea
\cite{Nreviews}. 
However we still
do not have a quantitative control of the SU($N=\infty$) 
theory and the phenomenology needs to assume, for example, 
that it is confining and, of course, `close to' SU(3).  
Lattice simulations can attempt to answer such questions
directly (albeit never exactly) and there has been substantial
progress in doing so this last decade, first in D=2+1 dimensions
\cite{mtNd3},
which I do not discuss here, and then in the physically
interesting case of D=3+1. (For a review of work in the earlier 
80's, when lattice calculations were not yet precise enough
to be so useful, see e.g.
\cite{Dasrev}.) Here I focus on what these `modern' lattice 
calculations teach us about the properties of SU($N$) gauge
theories at large $N$. I will begin with some motivation for 
these calculations.

A gluon loop on a gluon propagator comes with a factor of $g^2 N$. 
One easily sees that $g^2$ is in fact the smallest power 
of the coupling that comes with a factor of $N$. So if one wants 
an $N\to \infty$ limit that is not given by either a free
field theory or by infinite order diagrams on all length scales
(in neither case would we get something like QCD),
then one needs to take the limit keeping $g^2 N$ fixed
\cite{thooftN}. 
In Section~\ref{section_Coupling} we shall see that this 
standard all-order perturbative statement also appears to hold 
non-perturbatively. Using 't Hooft's double-line notation for 
gluons, diagrams can be categorised as lying on surfaces
of different topology, with more handles
corresponding to higher powers of $1/N$, so that   
in the $N=\infty$ limit only planar diagrams survive
\cite{thooftN}.
As the coupling $g^2 N$ becomes strong, the vertices of the diagram 
fill the surface more densely, defining a world-sheet of the kind 
one might expect in a string theory. This suggests that perhaps as
$N\to\infty$ the gauge theory can be described as a weakly
interacting string theory
\cite{thooftN}.
If the theory is, as one expects, linearly confining -- and in 
Section~\ref{section_Confinement}
we present evidence that it is -- then the confining flux tube
will behave like a string at long distances, described by some
effective string theory. At large $N$ this will presumably
coincide with the string theory that describes the SU($N$) 
gauge theory. In Section~\ref{section_Confinement}
we shall discuss what a study of the string spectrum teaches
us about this string theory. These `old' string theory ideas
\cite{Polybook}
have been recently complemented by the realisation 
that at $g^2N\to\infty$ and $N\to\infty$ gauge theories 
have a dual string description that is analytically tractable
\cite{AdS}. 
Determining the effective string theory numerically should
provide useful hints about what the dual theory might be 
in the physical weak-coupling limit, $g^2N\to 0$. 

In a confining theory there are no decays at $N=\infty$. This
is in contrast to what would happen in a  
non-confining theory where a coloured state could have a finite
decay width into other coloured  states. But once we constrain
states to be colour singlet we reduce the density of final states 
by factors of $N$ so that all decay widths vanish. In addition
there is no scattering between the colour singlet states. 
Think of two propagating mesons. A meson propagator is like a 
closed quark loop. Exchange two gluons between these two
closed loops and you clearly gain (up to) a factor of $N$,
but at the cost of $g^4 \propto 1/N^2$. So, no scattering
at $N=\infty$. However it is also easy to see that within a single
closed loop, planar interactions give factors of $g^2N$,
are not suppressed, and so 
there are non-trivial bound states. So we have what looks like 
a free theory, but it has a complex bound state spectrum
and so is non-trivial. If one is going to find room in D=3+1 for 
notions of e.g. integrability, it is here in the $N=\infty$ limit 
that they might find a suitable home.

Is this theory with no decays and no scattering similar
to the observed world of the strong interactions? First
we need the SU($\infty$) theory to be linearly confining -- and 
we provide evidence for this in Section~\ref{section_Confinement}.
Now we can ask: is this confining theory close to SU(3)? In
Section~\ref{section_Spectrum} we calculate the
lightest masses in the spectrum for several values of $N$,
and find that the SU(3) mass spectrum is indeed very close
to the (extrapolated) spectrum at $N=\infty$. This provides
support for the phenomenological relevance of the SU($\infty$) 
theory. And this provides motivation for trying to understand
that theory much better. In Section~\ref{section_Confinement}
we also see that the effective string theory describing long
confining flux tubes appears to be in the bosonic string
universlity class. More surprisingly, the energy of shorter
strings is close to the Nambu-Goto prediction and at smaller
$N$, where this question can be addressed, the string condensation
temperature is very close to that of the Nambu-Goto string action.
A new phenomenon for $N>3$ is the existence of new stable
strings. In Section~\ref{section_k-strings} we summarise the 
latest lattice calculations of the corresponding string tensions 
and find they lie between the `Casimir scaling' 
\cite{CS}
and `MQCD'
\cite{MQCD} 
conjectures. We remark how these new strings can contribute
to an $N$-dependence of the mass spectrum even in the
confining phase, contrary to naive expectations. In 
Section~\ref{section_Deconfinement} we learn how
the large-$N$ gauge theory deconfines. Contrary to some
speculations that the $N=\infty$ transition might be second order,
partly motivated by the weakness of the first order transition 
in SU(3), we will show it to be robustly first order. 

At $N=\infty$ the expectation value of a product of gauge
invariant operators factorises into the product of the
respective expectation values -- by the same argument that
there is no scattering. This suggests that a single
gauge orbit -- Witten's Master Field 
\cite{EW-MF}
-- dominates the Path Integral calculation of all the physics 
in the confined phase. Since the physics is translation 
invariant, so must the Master Field be (for gauge invariant
quantities). This suggests that all we need to know is the
field in an arbitrarily small region to know it everywhere
-- even on one point if that can be made precise by a suitable
regularisation. On the lattice this is achieved through
(twisted) Eguchi-Kawai reduction
\cite{EK-twist}.

In the deconfinement transition one sees explicitly how
the large-$N$ behaviour of various quantities -- latent
heat, interface tension, fluctuations -- means that a
`phase transition' occurs on ever smaller volumes
as $N\to\infty$. We shall also see, by a heuristic but
physical argument, why the imposition of twisted boundary
conditions is required to remain in the right phase.  
Precisely at the deconfining temperature, $T_c$, there 
is a different Master Field of the Euclidean Path Integral
for the confined and deconfined phases. Through hysteresis
this extends either side of this temperature; conceivably
to all $T$. Indeed the Euclidean system possesses $N$
different phases, and hence Master Fields in the deconfined
phase. This multiplication of master fields is not peculiar
to deconfinement. For example, intertwining $\theta$-vacua
\cite{EWtheta,MStheta} 
would lead to $N$ non-degenerate vacua at $\theta=0$ which 
become absolutely stable at $N=\infty$, as discussed in
Section~\ref{section_Topology}. Each of these vacua will 
have its corresponding master field. 

For $N\geq 5$ one finds, in the lattice gauge theory with the
standard Wilson plaquette action, a first order `bulk'
phase transition at a particular value of the inverse `t Hooft
coupling $\lambda_c(a) = g^2(a)N \simeq 1/3$. (We
write the bare coupling as a running coupling on the scale $a$.)
This is essentially the same as the $D=1+1$, $N=\infty$ Gross-Witten
phase transition
\cite{GW}. 
For $\lambda(a) > \lambda_c(a)$ the vacuum is non-perturbative 
on all lengths scales, so that the confining string tension is
$O(1)$ in lattice units. For  $\lambda(a) < \lambda_c(a)$ the
vacuum is perturbative on the shortest distance scales, and
hence asymptotically free as $a\to 0$, and the  string tension is 
$O(1)$ in physical units. This transition appears as a lattice 
peculiarity but, as has recently been discovered
\cite{HN04,HNRN},
there appears to be an analogous non-analyticity at $N=\infty$ that 
occurs as we increase the size of a Wilson loop: at a certain critical
size the eigenvalue spectrum of the loop changes non-analytically
\cite{HN04,HNRN}.
This is possible because at $N=\infty$ the number of physically relevant
degrees of freedom per unit volume is infinite.
It appears that this critical size is fixed in physical units and will
survive in the continuum limit.

There are, of course, other non-analyticities as $N\to\infty$. 
For example, we shall see in Section~\ref{section_Topology} that the
instanton size distribution exactly vanishes for sizes up to some 
critical size. However this can be understood as due to the
factor $\exp(-8\pi^2/g^2) \propto \exp(-cN)$ that dominates the weighting
of small instantons. We do not know of any such simple argument in the
case of Wilson loops. Indeed
we might conjecture that this non-analyticity provides an explanation
for the puzzlingly rapid  transition between short and long distance
physics that is observed experimentally
\cite{JaffeECT}; 
i.e. as soon as one is at values
of $Q^2$ where one can apply perturbation theory, one finds that there
is little room for the higher twist operators that one might expect
to parametrise the transition from perturbative to non-perturbative
physics -- `precocious scaling'. The non-analyticity discovered in
\cite{HN04,HNRN}
suggests that for $l < l_c$ we
can calculate the wilson loop perturbatively, while for $l > l_c$
it is confining and non-perturbative. At $N=\infty$ the transition
is infinitely sharp; in SU(3) and QCD it might become a very rapid
cross-over, explaining the phenomenon of precocious scaling.

The presence of such a `phase transition' as we increase the distance,
might effectively disconnect the confining theory 
from its short distance perturbative framework. In this disembodied 
confining theory the coupling is never small, it is confining on 
all available length scales and one never needs to discuss gluons. 
This raises, for example, the possibility
of a dual string theory in which the coupling need not be large.  
Such a dual theory might be analytically tractable. It also raises the
possibility that at $N=\infty$ the same confining theory can have different
ultraviolet completions. That is to say, to solve the theory we do not
necessarily need to solve the full non-Abelian gauge theory. These 
conjectures are speculative, of course, but they certainly provide 
motivation for clarifying
\cite{fbmthv}
the nature of this remarkable  non-analyticity.

\section{Lattice}

We will calculate Euclidean Feynman Path Integrals numerically. 
This requires a finite number of degrees of freedom, so we
discretise continuous space-time and make the volume finite
by going to a hypercubic lattice on a 4-torus. Since the theory 
is renormalisible and has a mass gap, the errors induced
by this should rapidly disappear as the lattice spacing is 
reduced and the volume enlarged. The lattice
spacing is $a$ and the size of the $\mu$-torus is 
$l_\mu=aL_\mu$. The degrees of freedom are SU($N$) matrices, 
$U_l$, defined on the links $l$ of the lattice. The
partition function is 
\begin{equation}
{\mathcal Z}(\beta)
=
\int \prod_l dU_l 
e^{-\beta \sum_p\{ 1 - \frac{1}{N}{\mathrm{ReTr}}u_p\}}
\ \ \ \ ; \ \ \ \
\beta=\frac{2N}{g^2}
\label{eqn_Zlattice} 
\end{equation} 
where $u_p$ is the ordered product of matrices around the
boundary of the elementary square (plaquette) labelled by $p$
and $g^2$ is the bare coupling. This is the standard
Wilson plaquette action and one can easily see that for
smooth fields it reduce to the usual continuum gauge theory.
Since the theory is asymptotically free and since the bare coupling 
is a running coupling on length scale $a$, the continuum limit is 
approached by tuning $\beta = 2N/g^2(a) \to \infty$. 
As we remarked earlier, one expects from the diagrammatic
analysis that for large $N$ the value of $a$ is fixed  
in physical units (e.g. in units of the mass gap) if one
keeps the 't Hooft coupling $\lambda(a) \equiv g^2(a)N$
fixed i.e. $\beta \propto N^2$. This will be confirmed below.

The lattice path integral in eqn(\ref{eqn_Zlattice}) is no easier
to calculate analytically than the original continuum 
version. However because the number of integrations
is now finite, we can attempt a numerical evaluation.
The number of integrations is large and so the 
natural method to use is the (Markovian) Monte Carlo.
The Monte Carlo generates `points' in the integration
space. Each such `point' is an explicit lattice gauge
field i.e. an SU($N$) matrix on every link of the lattice.
These fields are generated with the measure
\be
{\mathcal D}U = \prod_l dU_l 
e^{-\beta \sum_p\{ 1 - \frac{1}{N}{\mathrm{ReTr}}u_p\}}
\label{eqn_measMC}
\ee
so if we generate $n_c$ such `points', i.e.
$\{U_{\mu}(n) ; \mu=0,..,3 ; n=1,...,L^4\}^I \ ; \ I=1,...,n_c$,
then the expectation value of $\Psi$ will be just the average
over these fields:
\be
\langle \Psi_L(U) \rangle
=
{\frac{1}{n_c}} \sum_{I=1}^{n_c} \Psi_L(U^I)
\pm
O(\frac{1}{\sqrt{n_c}}).
\label{eqn_avMC}
\ee
I have made explicit here the statistical error which
decreases as the square root of the number of field
configurations -- as one would expect for such 
a probabilistic estimate.

We calculate masses from Euclidean correlation functions
\begin{equation}
C(t=an_t) 
\equiv \langle \phi^\dagger (t) \phi(0) \rangle
=
\sum_{n}  |\langle n |\phi(0) | vac \rangle|^2 e^{-aE_n n_t}
\ \ \ \ ; \ \ \ \
E_i \leq E_{i+1}
\label{eqn_corrln} 
\end{equation}
which we evaluate numerically as just described.
Note that all energies will be obtained in lattice units,
$aE_n$. At large $n_t$ the lightest state will dominate
$C(t=an_t)$ and can be easily extracted. Unfortunately
the statistical error in the calculation of eqn(\ref{eqn_avMC}),
is more-or-less independent of $n_t$, since the average fluctuation
squared around the correlator is itself a higher-order correlator 
which, one can easily verify, has a disconnected piece. Thus 
the error to signal ratio grows exponentially with $n_t$
and one needs $C(t=an_t)$ to be dominated by the lightest state
at small $n_t$ i.e. one needs  $\phi$ to be a good wave-functional
for the desired state. Standard techniques now exist to achieve
this, and can be used within a variational calculation, based
on the $\exp(-aH)$ implicit in $C(t=an_t)$, to obtain excited
as well as ground state energies
\cite{blmtuw-glue04}. 
However it should be apparent
that the larger the energy, the less accurate the calculation.

To calculate glueball masses we use operators that are based on 
contractible Wilson loops.
We calculate the string tension from the energy of the 
lightest flux loop that winds around a spatial torus,
and use operators based on a Wilson line that encircles
the torus.
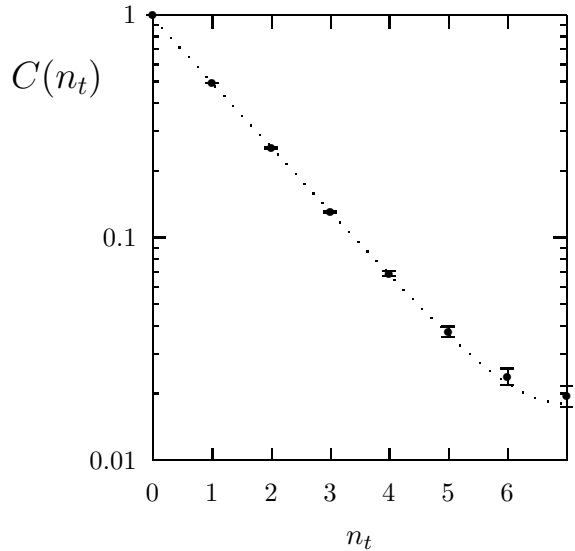
\begin	{figure}[ht]
\begin	{center}
\leavevmode
\setlength{\unitlength}{0.240900pt}
\ifx\plotpoint\undefined\newsavebox{\plotpoint}\fi
\sbox{\plotpoint}{\rule[-0.200pt]{0.400pt}{0.400pt}}%
\begin{picture}(1125,900)(0,0)
\font\gnuplot=cmr10 at 12pt
\gnuplot
\sbox{\plotpoint}{\rule[-0.200pt]{0.400pt}{0.400pt}}%
\put(400.0,150.0){\rule[-0.200pt]{4.818pt}{0.400pt}}
\put(375,150){\makebox(0,0)[r]{ \ \ {$0.01$}}}
\put(1030.0,150.0){\rule[-0.200pt]{4.818pt}{0.400pt}}
\put(400.0,255.0){\rule[-0.200pt]{2.409pt}{0.400pt}}
\put(1040.0,255.0){\rule[-0.200pt]{2.409pt}{0.400pt}}
\put(400.0,317.0){\rule[-0.200pt]{2.409pt}{0.400pt}}
\put(1040.0,317.0){\rule[-0.200pt]{2.409pt}{0.400pt}}
\put(400.0,361.0){\rule[-0.200pt]{2.409pt}{0.400pt}}
\put(1040.0,361.0){\rule[-0.200pt]{2.409pt}{0.400pt}}
\put(400.0,395.0){\rule[-0.200pt]{2.409pt}{0.400pt}}
\put(1040.0,395.0){\rule[-0.200pt]{2.409pt}{0.400pt}}
\put(400.0,422.0){\rule[-0.200pt]{2.409pt}{0.400pt}}
\put(1040.0,422.0){\rule[-0.200pt]{2.409pt}{0.400pt}}
\put(400.0,446.0){\rule[-0.200pt]{2.409pt}{0.400pt}}
\put(1040.0,446.0){\rule[-0.200pt]{2.409pt}{0.400pt}}
\put(400.0,466.0){\rule[-0.200pt]{2.409pt}{0.400pt}}
\put(1040.0,466.0){\rule[-0.200pt]{2.409pt}{0.400pt}}
\put(400.0,484.0){\rule[-0.200pt]{2.409pt}{0.400pt}}
\put(1040.0,484.0){\rule[-0.200pt]{2.409pt}{0.400pt}}
\put(400.0,500.0){\rule[-0.200pt]{4.818pt}{0.400pt}}
\put(375,500){\makebox(0,0)[r]{ \ \ {$0.1$}}}
\put(1030.0,500.0){\rule[-0.200pt]{4.818pt}{0.400pt}}
\put(400.0,605.0){\rule[-0.200pt]{2.409pt}{0.400pt}}
\put(1040.0,605.0){\rule[-0.200pt]{2.409pt}{0.400pt}}
\put(400.0,667.0){\rule[-0.200pt]{2.409pt}{0.400pt}}
\put(1040.0,667.0){\rule[-0.200pt]{2.409pt}{0.400pt}}
\put(400.0,711.0){\rule[-0.200pt]{2.409pt}{0.400pt}}
\put(1040.0,711.0){\rule[-0.200pt]{2.409pt}{0.400pt}}
\put(400.0,745.0){\rule[-0.200pt]{2.409pt}{0.400pt}}
\put(1040.0,745.0){\rule[-0.200pt]{2.409pt}{0.400pt}}
\put(400.0,772.0){\rule[-0.200pt]{2.409pt}{0.400pt}}
\put(1040.0,772.0){\rule[-0.200pt]{2.409pt}{0.400pt}}
\put(400.0,796.0){\rule[-0.200pt]{2.409pt}{0.400pt}}
\put(1040.0,796.0){\rule[-0.200pt]{2.409pt}{0.400pt}}
\put(400.0,816.0){\rule[-0.200pt]{2.409pt}{0.400pt}}
\put(1040.0,816.0){\rule[-0.200pt]{2.409pt}{0.400pt}}
\put(400.0,834.0){\rule[-0.200pt]{2.409pt}{0.400pt}}
\put(1040.0,834.0){\rule[-0.200pt]{2.409pt}{0.400pt}}
\put(400.0,850.0){\rule[-0.200pt]{4.818pt}{0.400pt}}
\put(375,850){\makebox(0,0)[r]{ \ \ {$1$}}}
\put(1030.0,850.0){\rule[-0.200pt]{4.818pt}{0.400pt}}
\put(400.0,150.0){\rule[-0.200pt]{0.400pt}{4.818pt}}
\put(400,100){\makebox(0,0){$0$}}
\put(400.0,830.0){\rule[-0.200pt]{0.400pt}{4.818pt}}
\put(493.0,150.0){\rule[-0.200pt]{0.400pt}{4.818pt}}
\put(493,100){\makebox(0,0){$1$}}
\put(493.0,830.0){\rule[-0.200pt]{0.400pt}{4.818pt}}
\put(586.0,150.0){\rule[-0.200pt]{0.400pt}{4.818pt}}
\put(586,100){\makebox(0,0){$2$}}
\put(586.0,830.0){\rule[-0.200pt]{0.400pt}{4.818pt}}
\put(679.0,150.0){\rule[-0.200pt]{0.400pt}{4.818pt}}
\put(679,100){\makebox(0,0){$3$}}
\put(679.0,830.0){\rule[-0.200pt]{0.400pt}{4.818pt}}
\put(771.0,150.0){\rule[-0.200pt]{0.400pt}{4.818pt}}
\put(771,100){\makebox(0,0){$4$}}
\put(771.0,830.0){\rule[-0.200pt]{0.400pt}{4.818pt}}
\put(864.0,150.0){\rule[-0.200pt]{0.400pt}{4.818pt}}
\put(864,100){\makebox(0,0){$5$}}
\put(864.0,830.0){\rule[-0.200pt]{0.400pt}{4.818pt}}
\put(957.0,150.0){\rule[-0.200pt]{0.400pt}{4.818pt}}
\put(957,100){\makebox(0,0){$6$}}
\put(957.0,830.0){\rule[-0.200pt]{0.400pt}{4.818pt}}
\put(400.0,150.0){\rule[-0.200pt]{156.585pt}{0.400pt}}
\put(1050.0,150.0){\rule[-0.200pt]{0.400pt}{168.630pt}}
\put(400.0,850.0){\rule[-0.200pt]{156.585pt}{0.400pt}}
\put(250,750){\makebox(0,0){\Large{$C(n_t)$}}}
\put(725,25){\makebox(0,0){\large{$n_t$}}}
\put(400.0,150.0){\rule[-0.200pt]{0.400pt}{168.630pt}}
\put(400,850){\usebox{\plotpoint}}
\put(390.0,850.0){\rule[-0.200pt]{4.818pt}{0.400pt}}
\put(390.0,850.0){\rule[-0.200pt]{4.818pt}{0.400pt}}
\put(493,743){\usebox{\plotpoint}}
\put(483.0,743.0){\rule[-0.200pt]{4.818pt}{0.400pt}}
\put(483.0,743.0){\rule[-0.200pt]{4.818pt}{0.400pt}}
\put(586.0,640.0){\rule[-0.200pt]{0.400pt}{0.482pt}}
\put(576.0,640.0){\rule[-0.200pt]{4.818pt}{0.400pt}}
\put(576.0,642.0){\rule[-0.200pt]{4.818pt}{0.400pt}}
\put(679.0,538.0){\rule[-0.200pt]{0.400pt}{0.964pt}}
\put(669.0,538.0){\rule[-0.200pt]{4.818pt}{0.400pt}}
\put(669.0,542.0){\rule[-0.200pt]{4.818pt}{0.400pt}}
\put(771.0,440.0){\rule[-0.200pt]{0.400pt}{1.686pt}}
\put(761.0,440.0){\rule[-0.200pt]{4.818pt}{0.400pt}}
\put(761.0,447.0){\rule[-0.200pt]{4.818pt}{0.400pt}}
\put(864.0,344.0){\rule[-0.200pt]{0.400pt}{3.854pt}}
\put(854.0,344.0){\rule[-0.200pt]{4.818pt}{0.400pt}}
\put(854.0,360.0){\rule[-0.200pt]{4.818pt}{0.400pt}}
\put(957.0,268.0){\rule[-0.200pt]{0.400pt}{6.263pt}}
\put(947.0,268.0){\rule[-0.200pt]{4.818pt}{0.400pt}}
\put(947.0,294.0){\rule[-0.200pt]{4.818pt}{0.400pt}}
\put(1050.0,234.0){\rule[-0.200pt]{0.400pt}{7.950pt}}
\put(1040.0,234.0){\rule[-0.200pt]{4.818pt}{0.400pt}}
\put(400,850){\circle*{12}}
\put(493,743){\circle*{12}}
\put(586,641){\circle*{12}}
\put(679,540){\circle*{12}}
\put(771,443){\circle*{12}}
\put(864,352){\circle*{12}}
\put(957,282){\circle*{12}}
\put(1050,252){\circle*{12}}
\put(1040.0,267.0){\rule[-0.200pt]{4.818pt}{0.400pt}}
\put(400,845){\usebox{\plotpoint}}
\put(400.00,845.00){\usebox{\plotpoint}}
\put(413.64,829.36){\usebox{\plotpoint}}
\put(427.80,814.20){\usebox{\plotpoint}}
\put(441.75,798.86){\usebox{\plotpoint}}
\put(455.87,783.66){\usebox{\plotpoint}}
\put(469.93,768.41){\usebox{\plotpoint}}
\put(483.52,752.73){\usebox{\plotpoint}}
\put(497.59,737.48){\usebox{\plotpoint}}
\put(512.19,722.74){\usebox{\plotpoint}}
\put(525.77,707.11){\usebox{\plotpoint}}
\put(539.83,691.86){\usebox{\plotpoint}}
\put(553.63,676.37){\usebox{\plotpoint}}
\put(567.79,661.21){\usebox{\plotpoint}}
\put(581.94,646.06){\usebox{\plotpoint}}
\put(595.55,630.45){\usebox{\plotpoint}}
\put(609.73,615.32){\usebox{\plotpoint}}
\put(623.86,600.14){\usebox{\plotpoint}}
\put(637.47,584.53){\usebox{\plotpoint}}
\put(651.63,569.37){\usebox{\plotpoint}}
\put(666.04,554.45){\usebox{\plotpoint}}
\put(679.78,538.95){\usebox{\plotpoint}}
\put(693.66,523.56){\usebox{\plotpoint}}
\put(708.22,508.78){\usebox{\plotpoint}}
\put(722.35,493.59){\usebox{\plotpoint}}
\put(736.41,478.35){\usebox{\plotpoint}}
\put(750.69,463.31){\usebox{\plotpoint}}
\put(764.85,448.15){\usebox{\plotpoint}}
\put(779.01,432.99){\usebox{\plotpoint}}
\put(793.16,417.84){\usebox{\plotpoint}}
\put(807.80,403.20){\usebox{\plotpoint}}
\put(822.10,388.20){\usebox{\plotpoint}}
\put(836.86,373.69){\usebox{\plotpoint}}
\put(851.62,359.18){\usebox{\plotpoint}}
\put(866.42,344.64){\usebox{\plotpoint}}
\put(881.74,330.65){\usebox{\plotpoint}}
\put(897.05,316.67){\usebox{\plotpoint}}
\put(912.95,303.32){\usebox{\plotpoint}}
\put(928.93,290.19){\usebox{\plotpoint}}
\put(945.90,278.35){\usebox{\plotpoint}}
\put(963.26,267.00){\usebox{\plotpoint}}
\put(981.26,256.83){\usebox{\plotpoint}}
\put(1000.02,248.14){\usebox{\plotpoint}}
\put(1019.98,242.57){\usebox{\plotpoint}}
\put(1040.33,239.45){\usebox{\plotpoint}}
\put(1050,239){\usebox{\plotpoint}}
\end{picture}
\end	{center}
\vskip 0.025in
\caption{Correlation function where the lightest state
is a flux loop that winds around the spatial torus. Best
single energy cosh fit shown.}
\label{fig_corr}
\end 	{figure}
In Fig.~\ref{fig_corr} I show an example of the latter
\cite{hmmt-conf}. 
The calculation is in SU(6) on a $12\times 14^3$ lattice with
the flux loop winding around the $x$-torus. Shown also is the
best single exponential fit (actually a cosh because of the 
periodicity in $t$). It is clear that it dominates the correlator
from very small $t$ -- indeed the overlap of the operator on the
flux loop is $\simeq 0.97$. This is achieved by iterative smearing
of the fields and by a variational calculation (see e.g.
\cite{blmtuw-glue04}
for details).

\section{Confinement and Strings}
\label{section_Confinement}

Consider one spatial torus of size $l$ and all the other tori large.   
We calculate the mass of the lightest flux loop that winds once
around this torus. We expect 
\cite{string-corrn}
its energy to be
\begin{equation}
m(l) 
\stackrel{l\to\infty}{=}
\sigma l - \frac{\pi(d-2)}{6} \frac{c}{l}
\label{eqn_luscher} 
\end{equation}
where $c$ (times the  dimensional factor $d-2$) is the central 
charge of the effective string 
theory that describes the long-distance properties of the
confining flux tube. 
\begin	{figure}[ht]
\begin	{center}
\leavevmode
\input	{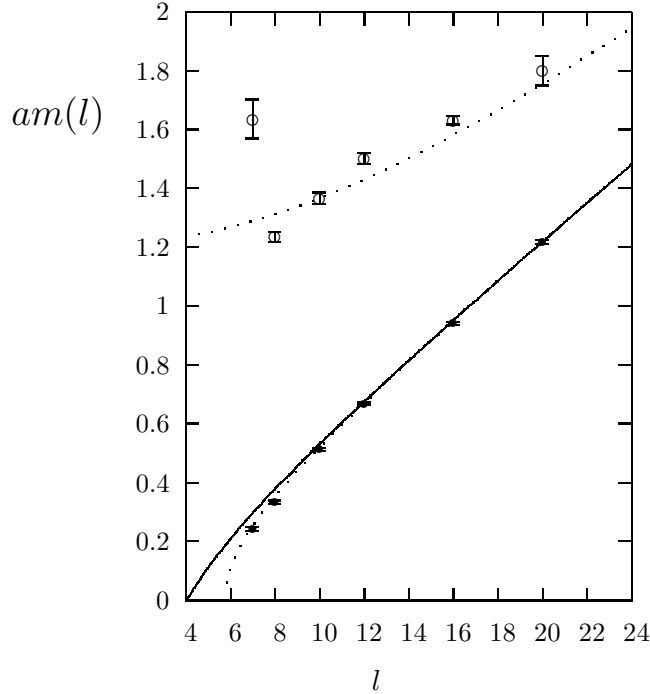}
\end	{center}
\vskip 0.025in
\caption{The masses of the lightest, $\bullet$, and first
excited, $\circ$, $k=1$ flux loops that wind
around a spatial torus of length $l$ in the SU(6) 
calculation at $\beta=25.05$. The dotted
lines are the predictions of the Nambu-Goto string action,
as in eqn(\ref{eqn_NGE}). The dynamical lower bound on the 
string length is $l_{min} = 1/aT_c \simeq 6.63$.}
\label{fig_K1}
\end 	{figure}
In Fig.~\ref{fig_K1} I show the results of a calculation of 
$a m(l)$ in SU(6) at a fixed value of $a$
\cite{hmmt-conf}.
We see linear confinement, and a fit with eqn(\ref{eqn_luscher})
to $l\geq 10, 12$ gives $c=1.16(8), 1.09(10)$ respectively.
This tells us that the effective string theory at long distances 
is a simple bosonic theory in the universality class of the
Nambu-Goto action. Indeed, if we fix the coefficient of the $1/l$ 
term to the bosonic string value $\pi/3$, then we find that we 
can obtain an acceptable fit to our whole range of $l$:
\begin{equation}
aE_0(l)
= 
0.06358(20) l - \frac{\pi}{3l} - \frac{19.4(3.2)}{l^3} 
\quad : \quad l \geq 7 \ \ , \chi^2/n_{df} = 0.9
\label{eqn_fitl4k1n6} 
\end{equation}
This is remarkable: since there is a 
minimum length for a periodic flux loop, which is 
$l_{min} = 1/aT_c \simeq 6.63$ in the present calculation,
the fit in eqn(\ref{eqn_fitl4k1n6}) essentially works all 
the way down to the shortest possible strings. (If we fit the
$O(1/l)$ term as well, then its coefficient comes to 0.94(16).)
Since the corrections in the pure gauge theory to $N=\infty$ 
are $O(1/N^2)$, we can assume that all this is also
true of the SU($N=\infty$) theory. In physical units the
lattice spacing is small, $a\surd\sigma \simeq 0.25$, so
we can assume that  this is also true of the continuum limit.
Finally, the longest string is $20a \simeq 5/\surd\sigma$
so we can assume we are seeing the asymptotic behaviour of a 
long string.  

In the Nambu-Goto string theory the spectrum is given by
\cite{arvis-string83,luscher-string04}
\begin{equation}
E_n(l) 
 = 
\sigma l 
\Bigl\{ 
1 + \frac{8\pi}{\sigma l^2}\bigl( n -  \frac{d-2}{24}\bigr)
\Bigr\}^\frac{1}{2}
\label{eqn_NGE} 
\end{equation}
In Fig.~\ref{fig_K1} we show that the one parameter fit with 
$E_0(l)$ to the lightest string mass works not too badly all the
way down to the minimum possible string length, $l_c=1/T_c$.
The $\chi^2$ is too large to be acceptable, but is much
smaller than for SU(4). This leaves open the intriguing 
(and unexpected) possibility that the 
Nambu-Goto string action describes confining strings on
all length scales at $N=\infty$.

\section{Spectrum} 
\label{section_Spectrum}

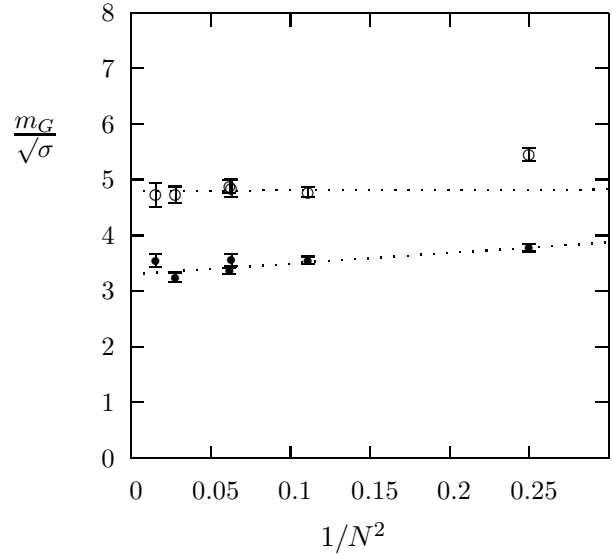
\begin	{figure}[ht]
\begin	{center}
\leavevmode
\setlength{\unitlength}{0.240900pt}
\ifx\plotpoint\undefined\newsavebox{\plotpoint}\fi
\sbox{\plotpoint}{\rule[-0.200pt]{0.400pt}{0.400pt}}%
\begin{picture}(1125,900)(0,0)
\font\gnuplot=cmr10 at 12pt
\gnuplot
\sbox{\plotpoint}{\rule[-0.200pt]{0.400pt}{0.400pt}}%
\put(300.0,150.0){\rule[-0.200pt]{4.818pt}{0.400pt}}
\put(275,150){\makebox(0,0)[r]{\ \ {$0$}}}
\put(1030.0,150.0){\rule[-0.200pt]{4.818pt}{0.400pt}}
\put(300.0,238.0){\rule[-0.200pt]{4.818pt}{0.400pt}}
\put(275,238){\makebox(0,0)[r]{\ \ {$1$}}}
\put(1030.0,238.0){\rule[-0.200pt]{4.818pt}{0.400pt}}
\put(300.0,325.0){\rule[-0.200pt]{4.818pt}{0.400pt}}
\put(275,325){\makebox(0,0)[r]{\ \ {$2$}}}
\put(1030.0,325.0){\rule[-0.200pt]{4.818pt}{0.400pt}}
\put(300.0,413.0){\rule[-0.200pt]{4.818pt}{0.400pt}}
\put(275,413){\makebox(0,0)[r]{\ \ {$3$}}}
\put(1030.0,413.0){\rule[-0.200pt]{4.818pt}{0.400pt}}
\put(300.0,500.0){\rule[-0.200pt]{4.818pt}{0.400pt}}
\put(275,500){\makebox(0,0)[r]{\ \ {$4$}}}
\put(1030.0,500.0){\rule[-0.200pt]{4.818pt}{0.400pt}}
\put(300.0,588.0){\rule[-0.200pt]{4.818pt}{0.400pt}}
\put(275,588){\makebox(0,0)[r]{\ \ {$5$}}}
\put(1030.0,588.0){\rule[-0.200pt]{4.818pt}{0.400pt}}
\put(300.0,675.0){\rule[-0.200pt]{4.818pt}{0.400pt}}
\put(275,675){\makebox(0,0)[r]{\ \ {$6$}}}
\put(1030.0,675.0){\rule[-0.200pt]{4.818pt}{0.400pt}}
\put(300.0,763.0){\rule[-0.200pt]{4.818pt}{0.400pt}}
\put(275,763){\makebox(0,0)[r]{\ \ {$7$}}}
\put(1030.0,763.0){\rule[-0.200pt]{4.818pt}{0.400pt}}
\put(300.0,850.0){\rule[-0.200pt]{4.818pt}{0.400pt}}
\put(275,850){\makebox(0,0)[r]{\ \ {$8$}}}
\put(1030.0,850.0){\rule[-0.200pt]{4.818pt}{0.400pt}}
\put(300.0,150.0){\rule[-0.200pt]{0.400pt}{4.818pt}}
\put(300,100){\makebox(0,0){\ {$0$}}}
\put(300.0,830.0){\rule[-0.200pt]{0.400pt}{4.818pt}}
\put(425.0,150.0){\rule[-0.200pt]{0.400pt}{4.818pt}}
\put(425,100){\makebox(0,0){\ {$0.05$}}}
\put(425.0,830.0){\rule[-0.200pt]{0.400pt}{4.818pt}}
\put(550.0,150.0){\rule[-0.200pt]{0.400pt}{4.818pt}}
\put(550,100){\makebox(0,0){\ {$0.1$}}}
\put(550.0,830.0){\rule[-0.200pt]{0.400pt}{4.818pt}}
\put(675.0,150.0){\rule[-0.200pt]{0.400pt}{4.818pt}}
\put(675,100){\makebox(0,0){\ {$0.15$}}}
\put(675.0,830.0){\rule[-0.200pt]{0.400pt}{4.818pt}}
\put(800.0,150.0){\rule[-0.200pt]{0.400pt}{4.818pt}}
\put(800,100){\makebox(0,0){\ {$0.2$}}}
\put(800.0,830.0){\rule[-0.200pt]{0.400pt}{4.818pt}}
\put(925.0,150.0){\rule[-0.200pt]{0.400pt}{4.818pt}}
\put(925,100){\makebox(0,0){\ {$0.25$}}}
\put(925.0,830.0){\rule[-0.200pt]{0.400pt}{4.818pt}}
\put(300.0,150.0){\rule[-0.200pt]{180.675pt}{0.400pt}}
\put(1050.0,150.0){\rule[-0.200pt]{0.400pt}{168.630pt}}
\put(300.0,850.0){\rule[-0.200pt]{180.675pt}{0.400pt}}
\put(150,650){\makebox(0,0){\Large{${{m_G}\over{\surd\sigma}}$}}}
\put(650,25){\makebox(0,0){\large{$1/N^2$}}}
\put(300.0,150.0){\rule[-0.200pt]{0.400pt}{168.630pt}}
\put(925.0,475.0){\rule[-0.200pt]{0.400pt}{2.891pt}}
\put(915.0,475.0){\rule[-0.200pt]{4.818pt}{0.400pt}}
\put(915.0,487.0){\rule[-0.200pt]{4.818pt}{0.400pt}}
\put(578.0,455.0){\rule[-0.200pt]{0.400pt}{2.891pt}}
\put(568.0,455.0){\rule[-0.200pt]{4.818pt}{0.400pt}}
\put(568.0,467.0){\rule[-0.200pt]{4.818pt}{0.400pt}}
\put(455.0,439.0){\rule[-0.200pt]{0.400pt}{2.409pt}}
\put(445.0,439.0){\rule[-0.200pt]{4.818pt}{0.400pt}}
\put(445.0,449.0){\rule[-0.200pt]{4.818pt}{0.400pt}}
\put(458.0,452.0){\rule[-0.200pt]{0.400pt}{4.577pt}}
\put(448.0,452.0){\rule[-0.200pt]{4.818pt}{0.400pt}}
\put(448.0,471.0){\rule[-0.200pt]{4.818pt}{0.400pt}}
\put(370.0,427.0){\rule[-0.200pt]{0.400pt}{3.613pt}}
\put(360.0,427.0){\rule[-0.200pt]{4.818pt}{0.400pt}}
\put(360.0,442.0){\rule[-0.200pt]{4.818pt}{0.400pt}}
\put(339.0,450.0){\rule[-0.200pt]{0.400pt}{5.059pt}}
\put(329.0,450.0){\rule[-0.200pt]{4.818pt}{0.400pt}}
\put(925,481){\circle*{12}}
\put(578,461){\circle*{12}}
\put(455,444){\circle*{12}}
\put(458,462){\circle*{12}}
\put(370,434){\circle*{12}}
\put(339,461){\circle*{12}}
\put(329.0,471.0){\rule[-0.200pt]{4.818pt}{0.400pt}}
\put(925.0,617.0){\rule[-0.200pt]{0.400pt}{4.818pt}}
\put(915.0,617.0){\rule[-0.200pt]{4.818pt}{0.400pt}}
\put(915.0,637.0){\rule[-0.200pt]{4.818pt}{0.400pt}}
\put(578.0,560.0){\rule[-0.200pt]{0.400pt}{3.854pt}}
\put(568.0,560.0){\rule[-0.200pt]{4.818pt}{0.400pt}}
\put(568.0,576.0){\rule[-0.200pt]{4.818pt}{0.400pt}}
\put(455.0,567.0){\rule[-0.200pt]{0.400pt}{4.818pt}}
\put(445.0,567.0){\rule[-0.200pt]{4.818pt}{0.400pt}}
\put(445.0,587.0){\rule[-0.200pt]{4.818pt}{0.400pt}}
\put(458.0,560.0){\rule[-0.200pt]{0.400pt}{6.745pt}}
\put(448.0,560.0){\rule[-0.200pt]{4.818pt}{0.400pt}}
\put(448.0,588.0){\rule[-0.200pt]{4.818pt}{0.400pt}}
\put(370.0,551.0){\rule[-0.200pt]{0.400pt}{6.263pt}}
\put(360.0,551.0){\rule[-0.200pt]{4.818pt}{0.400pt}}
\put(360.0,577.0){\rule[-0.200pt]{4.818pt}{0.400pt}}
\put(339.0,545.0){\rule[-0.200pt]{0.400pt}{9.154pt}}
\put(329.0,545.0){\rule[-0.200pt]{4.818pt}{0.400pt}}
\put(925,627){\circle{18}}
\put(578,568){\circle{18}}
\put(455,577){\circle{18}}
\put(458,574){\circle{18}}
\put(370,564){\circle{18}}
\put(339,564){\circle{18}}
\put(329.0,583.0){\rule[-0.200pt]{4.818pt}{0.400pt}}
\put(300,439){\usebox{\plotpoint}}
\put(300.00,439.00){\usebox{\plotpoint}}
\put(320.65,440.71){\usebox{\plotpoint}}
\put(341.32,442.00){\usebox{\plotpoint}}
\put(362.01,443.14){\usebox{\plotpoint}}
\put(382.63,444.95){\usebox{\plotpoint}}
\put(403.32,446.00){\usebox{\plotpoint}}
\put(423.99,447.37){\usebox{\plotpoint}}
\put(444.64,449.00){\usebox{\plotpoint}}
\put(465.32,450.00){\usebox{\plotpoint}}
\put(485.97,451.57){\usebox{\plotpoint}}
\put(506.63,453.00){\usebox{\plotpoint}}
\put(527.32,454.04){\usebox{\plotpoint}}
\put(547.97,455.75){\usebox{\plotpoint}}
\put(568.62,457.45){\usebox{\plotpoint}}
\put(589.28,459.00){\usebox{\plotpoint}}
\put(609.97,460.00){\usebox{\plotpoint}}
\put(630.61,461.66){\usebox{\plotpoint}}
\put(651.27,463.00){\usebox{\plotpoint}}
\put(671.96,464.12){\usebox{\plotpoint}}
\put(692.61,465.83){\usebox{\plotpoint}}
\put(713.28,467.00){\usebox{\plotpoint}}
\put(733.94,468.28){\usebox{\plotpoint}}
\put(754.59,469.95){\usebox{\plotpoint}}
\put(775.28,471.00){\usebox{\plotpoint}}
\put(795.94,472.49){\usebox{\plotpoint}}
\put(816.59,474.00){\usebox{\plotpoint}}
\put(837.28,475.00){\usebox{\plotpoint}}
\put(857.92,476.62){\usebox{\plotpoint}}
\put(878.59,478.00){\usebox{\plotpoint}}
\put(899.28,479.16){\usebox{\plotpoint}}
\put(919.92,480.85){\usebox{\plotpoint}}
\put(940.59,482.00){\usebox{\plotpoint}}
\put(961.27,483.28){\usebox{\plotpoint}}
\put(981.92,484.99){\usebox{\plotpoint}}
\put(1002.61,486.00){\usebox{\plotpoint}}
\put(1023.26,487.47){\usebox{\plotpoint}}
\put(1043.91,489.00){\usebox{\plotpoint}}
\put(1050,489){\usebox{\plotpoint}}
\put(300,570){\usebox{\plotpoint}}
\put(300.00,570.00){\usebox{\plotpoint}}
\put(320.76,570.00){\usebox{\plotpoint}}
\put(341.51,570.00){\usebox{\plotpoint}}
\put(362.27,570.00){\usebox{\plotpoint}}
\put(383.02,570.00){\usebox{\plotpoint}}
\put(403.78,570.00){\usebox{\plotpoint}}
\put(424.53,570.00){\usebox{\plotpoint}}
\put(445.29,570.00){\usebox{\plotpoint}}
\put(466.04,570.00){\usebox{\plotpoint}}
\put(486.80,570.00){\usebox{\plotpoint}}
\put(507.49,571.00){\usebox{\plotpoint}}
\put(528.25,571.00){\usebox{\plotpoint}}
\put(549.00,571.00){\usebox{\plotpoint}}
\put(569.76,571.00){\usebox{\plotpoint}}
\put(590.51,571.00){\usebox{\plotpoint}}
\put(611.27,571.00){\usebox{\plotpoint}}
\put(632.03,571.00){\usebox{\plotpoint}}
\put(652.78,571.00){\usebox{\plotpoint}}
\put(673.54,571.00){\usebox{\plotpoint}}
\put(694.29,571.00){\usebox{\plotpoint}}
\put(715.05,571.00){\usebox{\plotpoint}}
\put(735.80,571.00){\usebox{\plotpoint}}
\put(756.54,571.22){\usebox{\plotpoint}}
\put(777.24,572.00){\usebox{\plotpoint}}
\put(798.00,572.00){\usebox{\plotpoint}}
\put(818.75,572.00){\usebox{\plotpoint}}
\put(839.51,572.00){\usebox{\plotpoint}}
\put(860.26,572.00){\usebox{\plotpoint}}
\put(881.02,572.00){\usebox{\plotpoint}}
\put(901.78,572.00){\usebox{\plotpoint}}
\put(922.53,572.00){\usebox{\plotpoint}}
\put(943.29,572.00){\usebox{\plotpoint}}
\put(964.04,572.00){\usebox{\plotpoint}}
\put(984.80,572.00){\usebox{\plotpoint}}
\put(1005.55,572.00){\usebox{\plotpoint}}
\put(1026.25,572.89){\usebox{\plotpoint}}
\put(1046.99,573.00){\usebox{\plotpoint}}
\put(1050,573){\usebox{\plotpoint}}
\end{picture}
\end	{center}
\vskip 0.025in
\caption{The lightest $0^{++}$, $\bullet$, and $2^{++}$, $\circ$
glueball masses expressed in units of the fundamental string 
tension, in the continuum limit, plotted against $1/N^2$. 
Dotted lines are extrapolations to $N=\infty$.}
\label{fig_gkNwa}
\end 	{figure}

The lightest $J^{PC} = 0^{++}$ and  $2^{++}$ glueballs turn
out to be the lightest states in the $a=0, V=\infty$ SU($N$)
gauge theory. In Fig.~\ref{fig_gkNwa} I plot the ratios of
these masses to the (simultaneously calculated) string tension,
obtained after a continuum extrapolation of the lattice results
for each value of $N$
\cite{blmtuw-glue04}. 
We see that a modest $O(1/N^2)$ correction
suffices to fit the ratios for $N\geq 3$: for these quantities
SU(3) is indeed close to  SU($\infty$). 

This is for only two states of course. For a much more detailed 
comparison of SU(3) and SU(8) see
\cite{hmmt-pom}.

\section{Pomeron} 
\label{section_Pomeron}

In 
\cite{hmmt-pom}
you will also find estimates for higher spin
glueballs. This requires novel lattice techniques because
of the reduced cubic rotational symmetry of our lattice.
This calculation  enables us to ask if the Pomeron trajectory is
the leading glueball trajectory (ignoring mixing). 
\begin	{figure}[ht]
\begin	{center}
\leavevmode
\input	{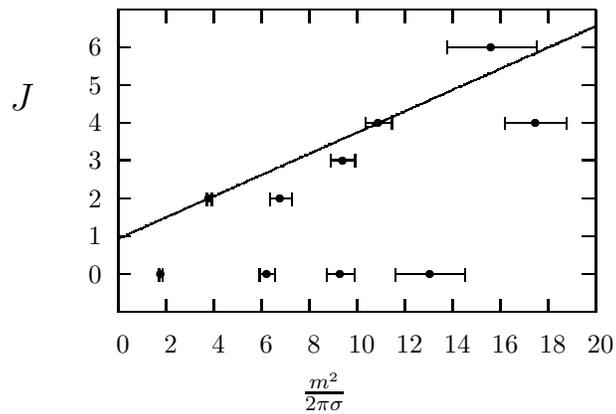}
\end	{center}
\vskip 0.025in
\caption{Chew-Frautschi plot of $PC=++$ states in the continuum 
SU(3) gauge theory. The leading Regge trajectory is shown.}
\label{fig_pom}
\end 	{figure}
Fig.~\ref{fig_pom} contains a  Chew-Frautschi plot
for the ${PC=++}$ sector of the D=3+1 SU(3) gauge theory.
Recall that $1/2\pi\sigma \simeq 1 {\mathrm GeV}$
is roughly the slope of the usual mesonic Regge trajectories.  
Assuming linear trajectories, the best fit to the intercept 
$\alpha_0$  and slope $\alpha^\prime$ of the leading trajectory is
\cite{hmmt-pom}
\begin{equation}
2\pi\sigma\alpha^\prime = 0.281(22) \quad , \quad 
\alpha_0 = 0.93(24)
\label{eqn_pom} 
\end{equation}
This provides quite convincing evidence for the old conjecture
that the Pomeron is in fact the leading glueball trajectory. It is
interesting to note that in 2+1 dimensions the leading
glueball trajectory has a very low interecept and so
is unimportant at high energies
\cite{hmmt-pom,hmmt-pomd3}. 
This is the case not only for SU(3) but for larger $N$ as well
\cite{hmmt-pom}.

\section{Coupling} 
\label{section_Coupling}

The diagrammatic expectation is that one keeps fixed the
`t Hooft coupling $\lambda = g^2N$ for a smooth large-$N$ 
limit. For theories with running couplings (as here)
this becomes 
\begin{equation}
\lambda(l\mu) = g^2(l\mu) N 
\stackrel{N\to\infty}{=}
{\mathrm{ind \ of}} \ N 
\label{eqn_ggN} 
\end{equation}
where the length scale $l$ is expressed in units of $\mu$,
which is some physical mass in the SU($N$) gauge theory
that one expects to have a smooth, non-zero large-$N$ limit.
We expect that if we plot $\lambda(l\mu)$ against $l\mu$
then as $N\to\infty$ all the curves for different $N$ will
lie on top of each other.
 
\begin	{figure}[ht]
\begin	{center}
\epsfig{figure=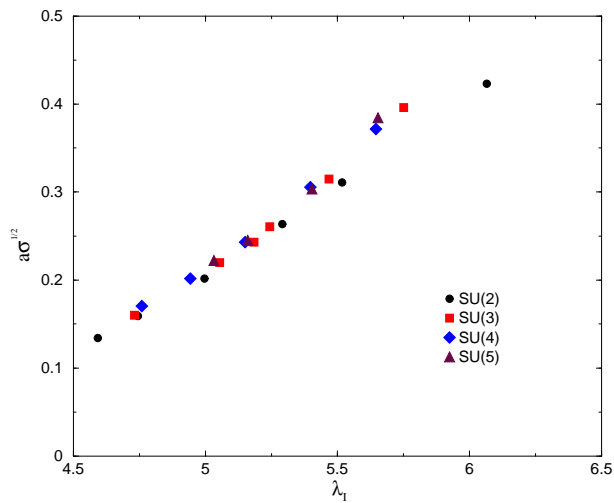, angle=270, width=8.0cm} 
\end	{center}
\vskip 0.05in
\caption{The square root of the string tension in lattice units,
$a\surd\sigma$, plotted against the 't Hooft coupling, 
$\lambda_I \equiv g^2_I N$.}
\label{fig_betaI}
\end 	{figure}

Let us choose $\mu=\surd\sigma$ and $l=a$. Then we can use
$\beta = 2N/g^2$ to define a coupling $g^2(a)$. This
is not ideal because of large lattice spacing corrections.
It is well known that the `mean field improved' coupling,
$\beta \langle \mathrm{ReTr} up\rangle/N = 2N/g_I^2(a)$, is much closer
to standard definitions. Defining 
$\lambda_I(a) = g_I^2(a) N$ we can plot its value against
the value of $a\surd\sigma$ obtained in the calculation with
the corrsponding $\beta$. We plot the results
\cite{blmt-glue01} 
in Fig.\ref{fig_betaI} for $2\leq N\leq 5$. This provides
nice non-perturbative confirmation of the usual perturbative
result.

\section{k-strings}
\label{section_k-strings}

Sources that transform as $\psi\to z^k\psi$ under a gauge
transformation of the centre, $z\in Z_N \subset SU(N)$, cannot 
be screened by gluons to a source that transforms otherwise.
Thus each sector $k$ has its own stable string tension, 
$\sigma_k$. Lattice calculations 
\cite{kstring}
have shown these are bound, $\sigma_k < k \sigma$, and
have focussed on a range of values that includes the
Casimir scaling 
\cite{CS}
and `MQCD'  conjectures
\cite{MQCD}.
Results from the most recent work on this topic 
\cite{blmtuw-glue04}
are listed in  Table~\ref{table_theoryK}. The 
results lie between the Casimir Scaling and MQCD
predictions.

\begin{table}
\begin{center}
\begin{tabular}{|c||c|c|c|}\hline
\multicolumn{4}{|c|}{ $\sigma_{k}/\sigma$ } \\ \hline
 (N,k) & Casimir scaling & this paper & `MQCD' \\ \hline
(4,2) & 1.333 & 1.370(20) &  1.414 \\
(4,2) & 1.333 & 1.358(33) &  1.414 \\
(6,2) & 1.600 & 1.675(31) &  1.732 \\
(6,3) & 1.800 & 1.886(61) &  2.000 \\
(8,2) & 1.714 & 1.779(51) &  1.848 \\
(8,3) & 2.143 & 2.38(10)  &  2.414 \\
(8,4) & 2.286 & 2.69(17)  &  2.613 \\ \hline
\end{tabular}
\caption{
Predictions of `Casimir Scaling' and `MQCD' compared against
our calculated values of
the ratio of the tension of the lightest $k$-string
to that of the fundamental ($k=1$) string. The second
SU(4) calculation is the one on the anisotropic lattice.}
\label{table_theoryK}
\end{center}
\end{table}

\begin	{figure}[ht]
\begin	{center}
\leavevmode
\setlength{\unitlength}{0.240900pt}
\ifx\plotpoint\undefined\newsavebox{\plotpoint}\fi
\sbox{\plotpoint}{\rule[-0.200pt]{0.400pt}{0.400pt}}%
\begin{picture}(1125,900)(0,0)
\font\gnuplot=cmr10 at 12pt
\gnuplot
\sbox{\plotpoint}{\rule[-0.200pt]{0.400pt}{0.400pt}}%
\put(350.0,150.0){\rule[-0.200pt]{4.818pt}{0.400pt}}
\put(325,150){\makebox(0,0)[r]{\ \ {$1.1$}}}
\put(1030.0,150.0){\rule[-0.200pt]{4.818pt}{0.400pt}}
\put(350.0,290.0){\rule[-0.200pt]{4.818pt}{0.400pt}}
\put(325,290){\makebox(0,0)[r]{\ \ {$1.2$}}}
\put(1030.0,290.0){\rule[-0.200pt]{4.818pt}{0.400pt}}
\put(350.0,430.0){\rule[-0.200pt]{4.818pt}{0.400pt}}
\put(325,430){\makebox(0,0)[r]{\ \ {$1.3$}}}
\put(1030.0,430.0){\rule[-0.200pt]{4.818pt}{0.400pt}}
\put(350.0,570.0){\rule[-0.200pt]{4.818pt}{0.400pt}}
\put(325,570){\makebox(0,0)[r]{\ \ {$1.4$}}}
\put(1030.0,570.0){\rule[-0.200pt]{4.818pt}{0.400pt}}
\put(350.0,710.0){\rule[-0.200pt]{4.818pt}{0.400pt}}
\put(325,710){\makebox(0,0)[r]{\ \ {$1.5$}}}
\put(1030.0,710.0){\rule[-0.200pt]{4.818pt}{0.400pt}}
\put(350.0,850.0){\rule[-0.200pt]{4.818pt}{0.400pt}}
\put(325,850){\makebox(0,0)[r]{\ \ {$1.6$}}}
\put(1030.0,850.0){\rule[-0.200pt]{4.818pt}{0.400pt}}
\put(350.0,150.0){\rule[-0.200pt]{0.400pt}{4.818pt}}
\put(350,100){\makebox(0,0){\ {$0$}}}
\put(350.0,830.0){\rule[-0.200pt]{0.400pt}{4.818pt}}
\put(467.0,150.0){\rule[-0.200pt]{0.400pt}{4.818pt}}
\put(467,100){\makebox(0,0){\ {$0.025$}}}
\put(467.0,830.0){\rule[-0.200pt]{0.400pt}{4.818pt}}
\put(583.0,150.0){\rule[-0.200pt]{0.400pt}{4.818pt}}
\put(583,100){\makebox(0,0){\ {$0.05$}}}
\put(583.0,830.0){\rule[-0.200pt]{0.400pt}{4.818pt}}
\put(700.0,150.0){\rule[-0.200pt]{0.400pt}{4.818pt}}
\put(700,100){\makebox(0,0){\ {$0.075$}}}
\put(700.0,830.0){\rule[-0.200pt]{0.400pt}{4.818pt}}
\put(817.0,150.0){\rule[-0.200pt]{0.400pt}{4.818pt}}
\put(817,100){\makebox(0,0){\ {$0.1$}}}
\put(817.0,830.0){\rule[-0.200pt]{0.400pt}{4.818pt}}
\put(933.0,150.0){\rule[-0.200pt]{0.400pt}{4.818pt}}
\put(933,100){\makebox(0,0){\ {$0.125$}}}
\put(933.0,830.0){\rule[-0.200pt]{0.400pt}{4.818pt}}
\put(1050.0,150.0){\rule[-0.200pt]{0.400pt}{4.818pt}}
\put(1050,100){\makebox(0,0){\ {$0.15$}}}
\put(1050.0,830.0){\rule[-0.200pt]{0.400pt}{4.818pt}}
\put(350.0,150.0){\rule[-0.200pt]{168.630pt}{0.400pt}}
\put(1050.0,150.0){\rule[-0.200pt]{0.400pt}{168.630pt}}
\put(350.0,850.0){\rule[-0.200pt]{168.630pt}{0.400pt}}
\put(150,650){\makebox(0,0){\Large{${{\sigma_{k=2}}\over{\sigma}}$}}}
\put(675,25){\makebox(0,0){\large{$a^2\sigma$}}}
\put(350.0,150.0){\rule[-0.200pt]{0.400pt}{168.630pt}}
\put(1002.0,479.0){\rule[-0.200pt]{0.400pt}{16.140pt}}
\put(992.0,479.0){\rule[-0.200pt]{4.818pt}{0.400pt}}
\put(992.0,546.0){\rule[-0.200pt]{4.818pt}{0.400pt}}
\put(844.0,519.0){\rule[-0.200pt]{0.400pt}{9.636pt}}
\put(834.0,519.0){\rule[-0.200pt]{4.818pt}{0.400pt}}
\put(834.0,559.0){\rule[-0.200pt]{4.818pt}{0.400pt}}
\put(763.0,452.0){\rule[-0.200pt]{0.400pt}{15.177pt}}
\put(753.0,452.0){\rule[-0.200pt]{4.818pt}{0.400pt}}
\put(753.0,515.0){\rule[-0.200pt]{4.818pt}{0.400pt}}
\put(692.0,534.0){\rule[-0.200pt]{0.400pt}{9.154pt}}
\put(682.0,534.0){\rule[-0.200pt]{4.818pt}{0.400pt}}
\put(682.0,572.0){\rule[-0.200pt]{4.818pt}{0.400pt}}
\put(634.0,486.0){\rule[-0.200pt]{0.400pt}{14.213pt}}
\put(624.0,486.0){\rule[-0.200pt]{4.818pt}{0.400pt}}
\put(624.0,545.0){\rule[-0.200pt]{4.818pt}{0.400pt}}
\put(534.0,504.0){\rule[-0.200pt]{0.400pt}{15.899pt}}
\put(524.0,504.0){\rule[-0.200pt]{4.818pt}{0.400pt}}
\put(524.0,570.0){\rule[-0.200pt]{4.818pt}{0.400pt}}
\put(459.0,469.0){\rule[-0.200pt]{0.400pt}{16.622pt}}
\put(449.0,469.0){\rule[-0.200pt]{4.818pt}{0.400pt}}
\put(1002,512){\circle*{12}}
\put(844,539){\circle*{12}}
\put(763,483){\circle*{12}}
\put(692,553){\circle*{12}}
\put(634,516){\circle*{12}}
\put(534,537){\circle*{12}}
\put(459,504){\circle*{12}}
\put(449.0,538.0){\rule[-0.200pt]{4.818pt}{0.400pt}}
\put(1006.0,497.0){\rule[-0.200pt]{0.400pt}{12.768pt}}
\put(996.0,497.0){\rule[-0.200pt]{4.818pt}{0.400pt}}
\put(996.0,550.0){\rule[-0.200pt]{4.818pt}{0.400pt}}
\put(801.0,536.0){\rule[-0.200pt]{0.400pt}{17.586pt}}
\put(791.0,536.0){\rule[-0.200pt]{4.818pt}{0.400pt}}
\put(791.0,609.0){\rule[-0.200pt]{4.818pt}{0.400pt}}
\put(654.0,482.0){\rule[-0.200pt]{0.400pt}{16.140pt}}
\put(644.0,482.0){\rule[-0.200pt]{4.818pt}{0.400pt}}
\put(644.0,549.0){\rule[-0.200pt]{4.818pt}{0.400pt}}
\put(532.0,465.0){\rule[-0.200pt]{0.400pt}{18.790pt}}
\put(522.0,465.0){\rule[-0.200pt]{4.818pt}{0.400pt}}
\put(1006,524){\circle{18}}
\put(801,573){\circle{18}}
\put(654,515){\circle{18}}
\put(532,504){\circle{18}}
\put(522.0,543.0){\rule[-0.200pt]{4.818pt}{0.400pt}}
\sbox{\plotpoint}{\rule[-0.500pt]{1.000pt}{1.000pt}}%
\put(350,528){\usebox{\plotpoint}}
\put(350.00,528.00){\usebox{\plotpoint}}
\put(370.76,528.00){\usebox{\plotpoint}}
\put(391.51,528.00){\usebox{\plotpoint}}
\put(412.27,528.00){\usebox{\plotpoint}}
\put(433.02,528.00){\usebox{\plotpoint}}
\put(453.78,528.00){\usebox{\plotpoint}}
\put(474.53,528.00){\usebox{\plotpoint}}
\put(495.29,528.00){\usebox{\plotpoint}}
\put(516.04,528.00){\usebox{\plotpoint}}
\put(536.73,529.00){\usebox{\plotpoint}}
\put(557.48,529.00){\usebox{\plotpoint}}
\put(578.24,529.00){\usebox{\plotpoint}}
\put(598.99,529.00){\usebox{\plotpoint}}
\put(619.75,529.00){\usebox{\plotpoint}}
\put(640.51,529.00){\usebox{\plotpoint}}
\put(661.26,529.00){\usebox{\plotpoint}}
\put(682.02,529.00){\usebox{\plotpoint}}
\put(702.77,529.00){\usebox{\plotpoint}}
\put(723.53,529.00){\usebox{\plotpoint}}
\put(744.28,529.00){\usebox{\plotpoint}}
\put(765.04,529.00){\usebox{\plotpoint}}
\put(785.79,529.00){\usebox{\plotpoint}}
\put(806.55,529.00){\usebox{\plotpoint}}
\put(827.31,529.00){\usebox{\plotpoint}}
\put(848.06,529.00){\usebox{\plotpoint}}
\put(868.79,529.40){\usebox{\plotpoint}}
\put(889.50,530.00){\usebox{\plotpoint}}
\put(910.26,530.00){\usebox{\plotpoint}}
\put(931.01,530.00){\usebox{\plotpoint}}
\put(951.77,530.00){\usebox{\plotpoint}}
\put(972.52,530.00){\usebox{\plotpoint}}
\put(993.28,530.00){\usebox{\plotpoint}}
\put(1014.03,530.00){\usebox{\plotpoint}}
\put(1034.79,530.00){\usebox{\plotpoint}}
\put(1050,530){\usebox{\plotpoint}}
\sbox{\plotpoint}{\rule[-0.200pt]{0.400pt}{0.400pt}}%
\put(350,512){\usebox{\plotpoint}}
\put(350.00,512.00){\usebox{\plotpoint}}
\put(370.68,513.00){\usebox{\plotpoint}}
\put(391.38,513.91){\usebox{\plotpoint}}
\put(412.12,514.00){\usebox{\plotpoint}}
\put(432.81,515.00){\usebox{\plotpoint}}
\put(453.49,516.00){\usebox{\plotpoint}}
\put(474.21,516.60){\usebox{\plotpoint}}
\put(494.93,517.00){\usebox{\plotpoint}}
\put(515.63,518.00){\usebox{\plotpoint}}
\put(536.31,519.00){\usebox{\plotpoint}}
\put(557.00,520.00){\usebox{\plotpoint}}
\put(577.73,520.25){\usebox{\plotpoint}}
\put(598.44,521.00){\usebox{\plotpoint}}
\put(619.12,522.00){\usebox{\plotpoint}}
\put(639.81,522.97){\usebox{\plotpoint}}
\put(660.56,523.00){\usebox{\plotpoint}}
\put(681.24,524.00){\usebox{\plotpoint}}
\put(701.93,525.00){\usebox{\plotpoint}}
\put(722.61,526.00){\usebox{\plotpoint}}
\put(743.32,526.62){\usebox{\plotpoint}}
\put(764.05,527.00){\usebox{\plotpoint}}
\put(784.74,528.00){\usebox{\plotpoint}}
\put(805.43,529.00){\usebox{\plotpoint}}
\put(826.16,529.31){\usebox{\plotpoint}}
\put(846.87,530.00){\usebox{\plotpoint}}
\put(867.55,531.00){\usebox{\plotpoint}}
\put(888.24,532.00){\usebox{\plotpoint}}
\put(908.92,532.99){\usebox{\plotpoint}}
\put(929.68,533.00){\usebox{\plotpoint}}
\put(950.36,534.00){\usebox{\plotpoint}}
\put(971.05,535.00){\usebox{\plotpoint}}
\put(991.75,535.82){\usebox{\plotpoint}}
\put(1012.49,536.00){\usebox{\plotpoint}}
\put(1033.17,537.00){\usebox{\plotpoint}}
\put(1050,538){\usebox{\plotpoint}}
\end{picture}
\end	{center}
\vskip 0.025in
\caption{Ratio of the tension of the $k=2$ string 
to that of the fundamental ($k=1$) string in SU(4), for the
anisotropic ($\circ$) and isotropic ($\bullet$) calculations. 
Plotted against the square of the (spatial) lattice spacing.
Continuum extrapolations are shown.}
\label{fig_kkn4}
\end 	{figure}
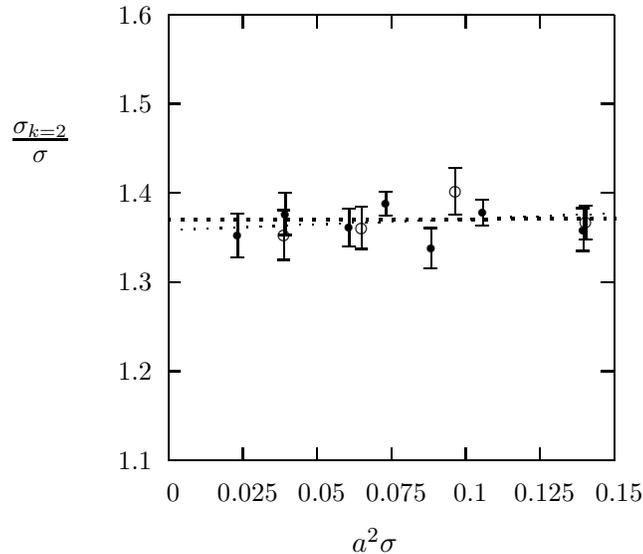

The values in Table~\ref{table_theoryK} are obtained after
an extrapolation to the continuum limit:
\begin{equation}
\frac{\sigma_k}{\sigma}(a) 
= 
\frac{\sigma_k}{\sigma}(0) 
+
c a^2\sigma
\label{eqn_contmk}
\end{equation}
from a range of $a$ where the leading correction proves sufficient.
(It is known that the leading correction for the plaquette
action is $O(a^2)$.) I show in Fig.~\ref{fig_kkn4} an example
\cite{blmtuw-glue04}
from SU(4), which superimposes two separate calculations.
It is clear that the continuum extrapolation is very well
determined. This is also true of the glueball calculations
discussed earlier.

All this has interesting implications for the mass spectrum.
One expects that (most) highly excited glueballs can
be described as closed loops of flux i.e. closed strings.
Normally one thinks of flux tubes in the fundamental 
representation. However closed loops of $k$-strings
will do just as well. These will scale with $\sigma_k$.
Thus as $N$ increases we expect the mass spectrum to
acquire new towers of states, based the stable strings
with $k\leq N/2$. These towers will be copies of each other
scaled up from the fundamental by  $\sqrt{\sigma_k/\sigma}$.
There is a hint of such a state in the comparison between
SU(3) and SU(8) in 
\cite{hmmt-pom}.
If $\sigma_k$ follows Casimir scaling then this will lead
to an `unexpected' $O(1/N)$ variation in some mass ratios.
It also implies that the `entropy' in the confined phase
is not $N$-independent, in contradiction to usual
assumptions (although this will not upset usual conclusions).

\section{Deconfinement and String Condensation}
\label{section_Deconfinement}

It is well-known that the SU(2) deconfining transition is
second order and that SU(3) is `weakly' first order.
Recent numerical studies of deconfinement in SU($N>3$)
gauge theories
\cite{blmtuwTc}
have shown that the transition becomes more strongly
first order as $N\uparrow$ and it is clear that it is
robustly first order in the $N=\infty$ limit. 

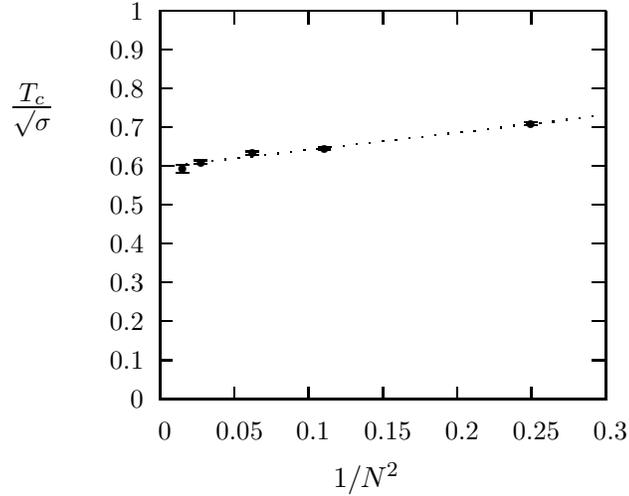
\begin	{figure}[ht]
\begin	{center}
\leavevmode
\setlength{\unitlength}{0.240900pt}
\ifx\plotpoint\undefined\newsavebox{\plotpoint}\fi
\sbox{\plotpoint}{\rule[-0.200pt]{0.400pt}{0.400pt}}%
\begin{picture}(1125,810)(0,0)
\font\gnuplot=cmr10 at 12pt
\gnuplot
\sbox{\plotpoint}{\rule[-0.200pt]{0.400pt}{0.400pt}}%
\put(350.0,150.0){\rule[-0.200pt]{4.818pt}{0.400pt}}
\put(325,150){\makebox(0,0)[r]{\ \ {$0$}}}
\put(1030.0,150.0){\rule[-0.200pt]{4.818pt}{0.400pt}}
\put(350.0,211.0){\rule[-0.200pt]{4.818pt}{0.400pt}}
\put(325,211){\makebox(0,0)[r]{\ \ {$0.1$}}}
\put(1030.0,211.0){\rule[-0.200pt]{4.818pt}{0.400pt}}
\put(350.0,272.0){\rule[-0.200pt]{4.818pt}{0.400pt}}
\put(325,272){\makebox(0,0)[r]{\ \ {$0.2$}}}
\put(1030.0,272.0){\rule[-0.200pt]{4.818pt}{0.400pt}}
\put(350.0,333.0){\rule[-0.200pt]{4.818pt}{0.400pt}}
\put(325,333){\makebox(0,0)[r]{\ \ {$0.3$}}}
\put(1030.0,333.0){\rule[-0.200pt]{4.818pt}{0.400pt}}
\put(350.0,394.0){\rule[-0.200pt]{4.818pt}{0.400pt}}
\put(325,394){\makebox(0,0)[r]{\ \ {$0.4$}}}
\put(1030.0,394.0){\rule[-0.200pt]{4.818pt}{0.400pt}}
\put(350.0,455.0){\rule[-0.200pt]{4.818pt}{0.400pt}}
\put(325,455){\makebox(0,0)[r]{\ \ {$0.5$}}}
\put(1030.0,455.0){\rule[-0.200pt]{4.818pt}{0.400pt}}
\put(350.0,516.0){\rule[-0.200pt]{4.818pt}{0.400pt}}
\put(325,516){\makebox(0,0)[r]{\ \ {$0.6$}}}
\put(1030.0,516.0){\rule[-0.200pt]{4.818pt}{0.400pt}}
\put(350.0,577.0){\rule[-0.200pt]{4.818pt}{0.400pt}}
\put(325,577){\makebox(0,0)[r]{\ \ {$0.7$}}}
\put(1030.0,577.0){\rule[-0.200pt]{4.818pt}{0.400pt}}
\put(350.0,638.0){\rule[-0.200pt]{4.818pt}{0.400pt}}
\put(325,638){\makebox(0,0)[r]{\ \ {$0.8$}}}
\put(1030.0,638.0){\rule[-0.200pt]{4.818pt}{0.400pt}}
\put(350.0,699.0){\rule[-0.200pt]{4.818pt}{0.400pt}}
\put(325,699){\makebox(0,0)[r]{\ \ {$0.9$}}}
\put(1030.0,699.0){\rule[-0.200pt]{4.818pt}{0.400pt}}
\put(350.0,760.0){\rule[-0.200pt]{4.818pt}{0.400pt}}
\put(325,760){\makebox(0,0)[r]{\ \ {$1$}}}
\put(1030.0,760.0){\rule[-0.200pt]{4.818pt}{0.400pt}}
\put(350.0,150.0){\rule[-0.200pt]{0.400pt}{4.818pt}}
\put(350,100){\makebox(0,0){\ {$0$}}}
\put(350.0,740.0){\rule[-0.200pt]{0.400pt}{4.818pt}}
\put(467.0,150.0){\rule[-0.200pt]{0.400pt}{4.818pt}}
\put(467,100){\makebox(0,0){\ {$0.05$}}}
\put(467.0,740.0){\rule[-0.200pt]{0.400pt}{4.818pt}}
\put(583.0,150.0){\rule[-0.200pt]{0.400pt}{4.818pt}}
\put(583,100){\makebox(0,0){\ {$0.1$}}}
\put(583.0,740.0){\rule[-0.200pt]{0.400pt}{4.818pt}}
\put(700.0,150.0){\rule[-0.200pt]{0.400pt}{4.818pt}}
\put(700,100){\makebox(0,0){\ {$0.15$}}}
\put(700.0,740.0){\rule[-0.200pt]{0.400pt}{4.818pt}}
\put(817.0,150.0){\rule[-0.200pt]{0.400pt}{4.818pt}}
\put(817,100){\makebox(0,0){\ {$0.2$}}}
\put(817.0,740.0){\rule[-0.200pt]{0.400pt}{4.818pt}}
\put(933.0,150.0){\rule[-0.200pt]{0.400pt}{4.818pt}}
\put(933,100){\makebox(0,0){\ {$0.25$}}}
\put(933.0,740.0){\rule[-0.200pt]{0.400pt}{4.818pt}}
\put(1050.0,150.0){\rule[-0.200pt]{0.400pt}{4.818pt}}
\put(1050,100){\makebox(0,0){\ {$0.3$}}}
\put(1050.0,740.0){\rule[-0.200pt]{0.400pt}{4.818pt}}
\put(350.0,150.0){\rule[-0.200pt]{168.630pt}{0.400pt}}
\put(1050.0,150.0){\rule[-0.200pt]{0.400pt}{146.949pt}}
\put(350.0,760.0){\rule[-0.200pt]{168.630pt}{0.400pt}}
\put(150,605){\makebox(0,0){\Large{${{T_c}\over{\surd\sigma}}$}}}
\put(675,25){\makebox(0,0){\large{$1/N^2$}}}
\put(350.0,150.0){\rule[-0.200pt]{0.400pt}{146.949pt}}
\put(933.0,580.0){\rule[-0.200pt]{0.400pt}{1.204pt}}
\put(923.0,580.0){\rule[-0.200pt]{4.818pt}{0.400pt}}
\put(923.0,585.0){\rule[-0.200pt]{4.818pt}{0.400pt}}
\put(609.0,542.0){\rule[-0.200pt]{0.400pt}{0.964pt}}
\put(599.0,542.0){\rule[-0.200pt]{4.818pt}{0.400pt}}
\put(599.0,546.0){\rule[-0.200pt]{4.818pt}{0.400pt}}
\put(496.0,534.0){\rule[-0.200pt]{0.400pt}{1.204pt}}
\put(486.0,534.0){\rule[-0.200pt]{4.818pt}{0.400pt}}
\put(486.0,539.0){\rule[-0.200pt]{4.818pt}{0.400pt}}
\put(415.0,519.0){\rule[-0.200pt]{0.400pt}{1.445pt}}
\put(405.0,519.0){\rule[-0.200pt]{4.818pt}{0.400pt}}
\put(405.0,525.0){\rule[-0.200pt]{4.818pt}{0.400pt}}
\put(386.0,505.0){\rule[-0.200pt]{0.400pt}{3.132pt}}
\put(376.0,505.0){\rule[-0.200pt]{4.818pt}{0.400pt}}
\put(933,583){\circle*{12}}
\put(609,544){\circle*{12}}
\put(496,537){\circle*{12}}
\put(415,522){\circle*{12}}
\put(386,512){\circle*{12}}
\put(376.0,518.0){\rule[-0.200pt]{4.818pt}{0.400pt}}
\put(350,514){\usebox{\plotpoint}}
\put(350.00,514.00){\usebox{\plotpoint}}
\put(370.55,516.94){\usebox{\plotpoint}}
\put(391.16,519.00){\usebox{\plotpoint}}
\put(411.73,521.68){\usebox{\plotpoint}}
\put(432.32,524.00){\usebox{\plotpoint}}
\put(452.89,526.56){\usebox{\plotpoint}}
\put(473.51,528.50){\usebox{\plotpoint}}
\put(494.06,531.44){\usebox{\plotpoint}}
\put(514.69,533.24){\usebox{\plotpoint}}
\put(535.24,536.18){\usebox{\plotpoint}}
\put(555.86,538.12){\usebox{\plotpoint}}
\put(576.41,541.06){\usebox{\plotpoint}}
\put(597.02,543.00){\usebox{\plotpoint}}
\put(617.59,545.80){\usebox{\plotpoint}}
\put(638.21,547.74){\usebox{\plotpoint}}
\put(658.75,550.68){\usebox{\plotpoint}}
\put(679.37,552.62){\usebox{\plotpoint}}
\put(699.93,555.49){\usebox{\plotpoint}}
\put(720.55,557.36){\usebox{\plotpoint}}
\put(741.10,560.30){\usebox{\plotpoint}}
\put(761.72,562.25){\usebox{\plotpoint}}
\put(782.26,565.18){\usebox{\plotpoint}}
\put(802.90,566.99){\usebox{\plotpoint}}
\put(823.45,569.92){\usebox{\plotpoint}}
\put(844.06,571.87){\usebox{\plotpoint}}
\put(864.61,574.80){\usebox{\plotpoint}}
\put(885.23,576.75){\usebox{\plotpoint}}
\put(905.79,579.54){\usebox{\plotpoint}}
\put(926.41,581.49){\usebox{\plotpoint}}
\put(946.96,584.42){\usebox{\plotpoint}}
\put(967.57,586.37){\usebox{\plotpoint}}
\put(988.12,589.30){\usebox{\plotpoint}}
\put(1008.75,591.11){\usebox{\plotpoint}}
\put(1029.30,594.04){\usebox{\plotpoint}}
\put(1049.91,595.99){\usebox{\plotpoint}}
\put(1050,596){\usebox{\plotpoint}}
\end{picture}
\end	{center}
\vskip 0.025in
\caption{The deconfining temperature in units of the
string tension for various SU($N$) gauge theories.
Large $N$ extrapolation shown.}
\label{fig_tcN}
\end 	{figure}

In Fig.~\ref{fig_tcN} I plot the value of the deconfining
temperature in units of the string tension, for $2\leq N\leq 8$.
We see that a modest $O(1/N^2)$ correction suffices to
fit these values. This is perhaps puzzling given that
SU(2) is second order. 

If nothing else happens as one increases $T$, one expects a 
string condensation deconfining transition to occur
\cite{poly-deconf}
simply because for strings of length $l$ the thermodynamic
suppression $\exp\{-E/T\} \sim \exp\{-\sigma l/T\}$ is
outweighed by the number of strings $\propto \exp\{+c_sl\}$
once $T > \sigma/c = T^s_c$, where we call $T^s_c$ the string 
condensation temperature. Thus any second order deconfining
transition may be driven by string condensation and the
value of $T^s_c/\sigma$ will then tell us something about
the effective string theory. For SU($N$) gauge theories
we have
\begin{equation}
\frac{T_c}{\surd\sigma} 
=
\begin{cases} 0.709(4) &:\quad SU(2), d=4\\
1.12(1)  &:\quad SU(2), d=3\\
0.98(2)  &:\quad SU(3), d=3
\end{cases}
\label{eqn_LatTc} 
\end{equation}
and it is interesting to compare it to the Nambu-Goto value
\begin{equation}
\frac{T^s_c}{\surd\sigma} 
=
\sqrt{\frac{3}{(d-2)\pi}}
\simeq \begin{cases} 0.691 &:\quad d=4\\
0.977  &:\quad d=3
\end{cases}
\label{eqn_NGTc} 
\end{equation}
obtained from the vanishing of $E_0(l)$ in eqn(\ref{eqn_NGE}).

As expected, one finds 
\cite{blmtuwTc}
that the latent heat $L_h \propto N^2$ and that the interface
tension, $\sigma_{cd}$, grows with $N$. It is easy to see
\cite{blmtuwTc}
that if $\sigma_{cd}\propto N$ then at $N=\infty$ the transition 
becomes infinitely sharp, i.e. a real phase transition, even
on a finite volume and there is no hysteresis. If 
$\sigma_{cd}\propto N^2$ then the hysteresis may be large --
perhaps even infinite! In that case, as we increase $T$ past
$T_c$ in the confined phase, we may continue to remain confined  
to a high enough $T$ that we encounter the string condensation
transition $T^s_c$. If so, then we can study string condensation
in the interesting $N=\infty$ limit.

Although one finds that one can calculate $T_c$ on ever smaller
volumes as $N\uparrow$, one also knows that once one of the spatial
sizes drops below $\sim 1/T_c$, flux loops that wind around that 
torus will condense, and this spontaneous symmetry breaking
will alter the thermal physics. This looks like an obstacle
to reducing the volume further as  $N\uparrow$. However one knows
how to prevent this -- twisted boundary conditions can force
a domain wall that restores the symmetry
\cite{walld3}.
For small sizes the wall unravels and the system sits in
the symmetric core.
Then one can continue to reduce $V$ with $N$. This provides
a physical interpretation of twisted Eguchi-Kawai.

\section{Topology, Instantons, $\theta$-vacua}
\label{section_Topology}

The topological susceptibility $\chi_t$ has a non-zero limit
at $N=\infty$ with a value that fits in with the 
Witten-Veneziano analysis of the $\eta^\prime$ mass
\cite{blmt-glue01,PisaQ}.
This is of course a non-leading effect 
\cite{OxfordTQ}
since $\langle Q^2\rangle = \langle Q\rangle^2 =0$
to leading order.

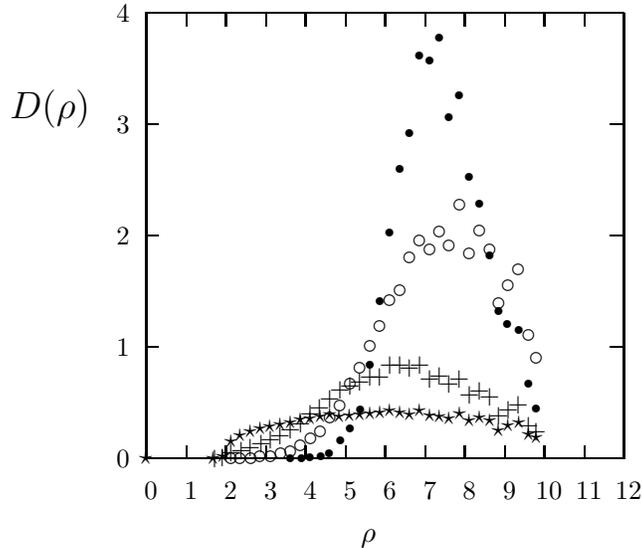
\begin	{figure}[ht]
\begin	{center}
\leavevmode
\setlength{\unitlength}{0.240900pt}
\ifx\plotpoint\undefined\newsavebox{\plotpoint}\fi
\sbox{\plotpoint}{\rule[-0.200pt]{0.400pt}{0.400pt}}%
\begin{picture}(1125,900)(0,0)
\font\gnuplot=cmr10 at 12pt
\gnuplot
\sbox{\plotpoint}{\rule[-0.200pt]{0.400pt}{0.400pt}}%
\put(300.0,150.0){\rule[-0.200pt]{4.818pt}{0.400pt}}
\put(275,150){\makebox(0,0)[r]{\ \ {$0$}}}
\put(1030.0,150.0){\rule[-0.200pt]{4.818pt}{0.400pt}}
\put(300.0,325.0){\rule[-0.200pt]{4.818pt}{0.400pt}}
\put(275,325){\makebox(0,0)[r]{\ \ {$1$}}}
\put(1030.0,325.0){\rule[-0.200pt]{4.818pt}{0.400pt}}
\put(300.0,500.0){\rule[-0.200pt]{4.818pt}{0.400pt}}
\put(275,500){\makebox(0,0)[r]{\ \ {$2$}}}
\put(1030.0,500.0){\rule[-0.200pt]{4.818pt}{0.400pt}}
\put(300.0,675.0){\rule[-0.200pt]{4.818pt}{0.400pt}}
\put(275,675){\makebox(0,0)[r]{\ \ {$3$}}}
\put(1030.0,675.0){\rule[-0.200pt]{4.818pt}{0.400pt}}
\put(300.0,850.0){\rule[-0.200pt]{4.818pt}{0.400pt}}
\put(275,850){\makebox(0,0)[r]{\ \ {$4$}}}
\put(1030.0,850.0){\rule[-0.200pt]{4.818pt}{0.400pt}}
\put(300.0,150.0){\rule[-0.200pt]{0.400pt}{4.818pt}}
\put(300,100){\makebox(0,0){\ {$0$}}}
\put(300.0,830.0){\rule[-0.200pt]{0.400pt}{4.818pt}}
\put(363.0,150.0){\rule[-0.200pt]{0.400pt}{4.818pt}}
\put(363,100){\makebox(0,0){\ {$1$}}}
\put(363.0,830.0){\rule[-0.200pt]{0.400pt}{4.818pt}}
\put(425.0,150.0){\rule[-0.200pt]{0.400pt}{4.818pt}}
\put(425,100){\makebox(0,0){\ {$2$}}}
\put(425.0,830.0){\rule[-0.200pt]{0.400pt}{4.818pt}}
\put(488.0,150.0){\rule[-0.200pt]{0.400pt}{4.818pt}}
\put(488,100){\makebox(0,0){\ {$3$}}}
\put(488.0,830.0){\rule[-0.200pt]{0.400pt}{4.818pt}}
\put(550.0,150.0){\rule[-0.200pt]{0.400pt}{4.818pt}}
\put(550,100){\makebox(0,0){\ {$4$}}}
\put(550.0,830.0){\rule[-0.200pt]{0.400pt}{4.818pt}}
\put(613.0,150.0){\rule[-0.200pt]{0.400pt}{4.818pt}}
\put(613,100){\makebox(0,0){\ {$5$}}}
\put(613.0,830.0){\rule[-0.200pt]{0.400pt}{4.818pt}}
\put(675.0,150.0){\rule[-0.200pt]{0.400pt}{4.818pt}}
\put(675,100){\makebox(0,0){\ {$6$}}}
\put(675.0,830.0){\rule[-0.200pt]{0.400pt}{4.818pt}}
\put(738.0,150.0){\rule[-0.200pt]{0.400pt}{4.818pt}}
\put(738,100){\makebox(0,0){\ {$7$}}}
\put(738.0,830.0){\rule[-0.200pt]{0.400pt}{4.818pt}}
\put(800.0,150.0){\rule[-0.200pt]{0.400pt}{4.818pt}}
\put(800,100){\makebox(0,0){\ {$8$}}}
\put(800.0,830.0){\rule[-0.200pt]{0.400pt}{4.818pt}}
\put(863.0,150.0){\rule[-0.200pt]{0.400pt}{4.818pt}}
\put(863,100){\makebox(0,0){\ {$9$}}}
\put(863.0,830.0){\rule[-0.200pt]{0.400pt}{4.818pt}}
\put(925.0,150.0){\rule[-0.200pt]{0.400pt}{4.818pt}}
\put(925,100){\makebox(0,0){\ {$10$}}}
\put(925.0,830.0){\rule[-0.200pt]{0.400pt}{4.818pt}}
\put(988.0,150.0){\rule[-0.200pt]{0.400pt}{4.818pt}}
\put(988,100){\makebox(0,0){\ {$11$}}}
\put(988.0,830.0){\rule[-0.200pt]{0.400pt}{4.818pt}}
\put(1050.0,150.0){\rule[-0.200pt]{0.400pt}{4.818pt}}
\put(1050,100){\makebox(0,0){\ {$12$}}}
\put(1050.0,830.0){\rule[-0.200pt]{0.400pt}{4.818pt}}
\put(300.0,150.0){\rule[-0.200pt]{180.675pt}{0.400pt}}
\put(1050.0,150.0){\rule[-0.200pt]{0.400pt}{168.630pt}}
\put(300.0,850.0){\rule[-0.200pt]{180.675pt}{0.400pt}}
\put(150,700){\makebox(0,0){\Large{$D(\rho)$}}}
\put(650,25){\makebox(0,0){\large{$\rho$}}}
\put(300.0,150.0){\rule[-0.200pt]{0.400pt}{168.630pt}}
\put(527,151){\circle*{12}}
\put(545,151){\circle*{12}}
\put(558,152){\circle*{12}}
\put(575,154){\circle*{12}}
\put(588,158){\circle*{12}}
\put(606,178){\circle*{12}}
\put(621,198){\circle*{12}}
\put(637,227){\circle*{12}}
\put(652,297){\circle*{12}}
\put(668,397){\circle*{12}}
\put(683,505){\circle*{12}}
\put(699,605){\circle*{12}}
\put(714,662){\circle*{12}}
\put(730,783){\circle*{12}}
\put(746,776){\circle*{12}}
\put(761,812){\circle*{12}}
\put(776,687){\circle*{12}}
\put(792,721){\circle*{12}}
\put(808,593){\circle*{12}}
\put(824,550){\circle*{12}}
\put(840,469){\circle*{12}}
\put(854,381){\circle*{12}}
\put(868,362){\circle*{12}}
\put(886,352){\circle*{12}}
\put(901,268){\circle*{12}}
\put(913,228){\circle*{12}}
\put(434,150){\circle{18}}
\put(448,151){\circle{18}}
\put(465,151){\circle{18}}
\put(479,154){\circle{18}}
\put(496,154){\circle{18}}
\put(512,158){\circle{18}}
\put(527,161){\circle{18}}
\put(543,171){\circle{18}}
\put(559,182){\circle{18}}
\put(574,192){\circle{18}}
\put(589,215){\circle{18}}
\put(605,234){\circle{18}}
\put(621,268){\circle{18}}
\put(636,292){\circle{18}}
\put(652,327){\circle{18}}
\put(667,359){\circle{18}}
\put(683,399){\circle{18}}
\put(699,414){\circle{18}}
\put(714,466){\circle{18}}
\put(730,493){\circle{18}}
\put(746,479){\circle{18}}
\put(761,507){\circle{18}}
\put(776,485){\circle{18}}
\put(793,549){\circle{18}}
\put(808,472){\circle{18}}
\put(824,508){\circle{18}}
\put(840,478){\circle{18}}
\put(854,395){\circle{18}}
\put(869,422){\circle{18}}
\put(885,447){\circle{18}}
\put(901,344){\circle{18}}
\put(913,309){\circle{18}}
\put(408,150){\makebox(0,0){$+$}}
\put(420,152){\makebox(0,0){$+$}}
\put(434,159){\makebox(0,0){$+$}}
\put(450,163){\makebox(0,0){$+$}}
\put(464,168){\makebox(0,0){$+$}}
\put(481,173){\makebox(0,0){$+$}}
\put(496,179){\makebox(0,0){$+$}}
\put(511,186){\makebox(0,0){$+$}}
\put(527,196){\makebox(0,0){$+$}}
\put(543,205){\makebox(0,0){$+$}}
\put(558,220){\makebox(0,0){$+$}}
\put(573,230){\makebox(0,0){$+$}}
\put(589,244){\makebox(0,0){$+$}}
\put(605,257){\makebox(0,0){$+$}}
\put(621,263){\makebox(0,0){$+$}}
\put(636,270){\makebox(0,0){$+$}}
\put(652,278){\makebox(0,0){$+$}}
\put(667,278){\makebox(0,0){$+$}}
\put(683,296){\makebox(0,0){$+$}}
\put(699,296){\makebox(0,0){$+$}}
\put(714,292){\makebox(0,0){$+$}}
\put(730,296){\makebox(0,0){$+$}}
\put(746,274){\makebox(0,0){$+$}}
\put(761,280){\makebox(0,0){$+$}}
\put(776,266){\makebox(0,0){$+$}}
\put(792,275){\makebox(0,0){$+$}}
\put(808,249){\makebox(0,0){$+$}}
\put(824,256){\makebox(0,0){$+$}}
\put(840,247){\makebox(0,0){$+$}}
\put(854,217){\makebox(0,0){$+$}}
\put(869,226){\makebox(0,0){$+$}}
\put(885,234){\makebox(0,0){$+$}}
\put(901,201){\makebox(0,0){$+$}}
\put(913,193){\makebox(0,0){$+$}}
\put(300,150){\makebox(0,0){$\star$}}
\put(406,150){\makebox(0,0){$\star$}}
\put(421,154){\makebox(0,0){$\star$}}
\put(434,177){\makebox(0,0){$\star$}}
\put(448,187){\makebox(0,0){$\star$}}
\put(464,193){\makebox(0,0){$\star$}}
\put(480,197){\makebox(0,0){$\star$}}
\put(495,201){\makebox(0,0){$\star$}}
\put(511,203){\makebox(0,0){$\star$}}
\put(527,207){\makebox(0,0){$\star$}}
\put(543,212){\makebox(0,0){$\star$}}
\put(558,213){\makebox(0,0){$\star$}}
\put(574,217){\makebox(0,0){$\star$}}
\put(589,219){\makebox(0,0){$\star$}}
\put(604,216){\makebox(0,0){$\star$}}
\put(620,218){\makebox(0,0){$\star$}}
\put(636,220){\makebox(0,0){$\star$}}
\put(652,221){\makebox(0,0){$\star$}}
\put(668,222){\makebox(0,0){$\star$}}
\put(683,225){\makebox(0,0){$\star$}}
\put(699,222){\makebox(0,0){$\star$}}
\put(714,219){\makebox(0,0){$\star$}}
\put(730,226){\makebox(0,0){$\star$}}
\put(746,218){\makebox(0,0){$\star$}}
\put(761,217){\makebox(0,0){$\star$}}
\put(776,213){\makebox(0,0){$\star$}}
\put(793,221){\makebox(0,0){$\star$}}
\put(808,210){\makebox(0,0){$\star$}}
\put(824,214){\makebox(0,0){$\star$}}
\put(840,210){\makebox(0,0){$\star$}}
\put(854,195){\makebox(0,0){$\star$}}
\put(869,202){\makebox(0,0){$\star$}}
\put(886,207){\makebox(0,0){$\star$}}
\put(901,188){\makebox(0,0){$\star$}}
\put(913,183){\makebox(0,0){$\star$}}
\end{picture}
\end	{center}
\vskip 0.025in
\caption{The instanton size density, $D(\rho)$, for
$N=2(\star),3(+),4(\circ),8(\bullet)$
on $16^4$ lattices with $a \simeq 1/8T_c$.}
\label{fig_drho16}
\end 	{figure}

In lattice calculations a very small instanton is a 
very big spike in the $F\tilde{F}$ density and is easy
to see. So in the instanton density plot 
\cite{OxfordTQ}
in Fig.~\ref{fig_drho16} the rapid suppression of small
instantons with $N$ is reliable -- and of course expected
from the 
\begin{equation}
D(\rho)
\stackrel{\rho\to 0}{\propto}
\frac{1}{\rho^5}
e^{-\frac{8\pi^2}{g^2(\rho)}} 
\propto 
\Bigl\{\frac{1}{\rho}\Bigr\}^{-(\frac{11N}{3}-5)}
\label{eqn_drho}
\end{equation}
factor that dominates the weight as $\rho\to 0$. This 
has consequences for the numerical calculation. Since
the Monte Carlo deforms the fields `continuously' (one
link matrix at a time) the only way to change $Q$ is to
shrink an (anti)instanton out of the lattice (or the reverse).
Thus the probability to change $Q$ is linked to the
probability to find an instanton of size 
$\rho \sim \mathrm{few} a$ and so is suppressed as
$\propto a^{11N/3-5}$. So for large $N$ a lattice calculation 
gets stuck in a fixed topological sector
\cite{blmt-glue01}.  
You could then try to calculate  $\langle Q^2\rangle$ from
a large sub-volume of a very large lattice volume.

Larger instantons are more ambiguous but Fig.~\ref{fig_drho16}
certainly suggests the interesting possibility
\begin{equation}
D(\rho) 
\stackrel{N\to\infty}{\longrightarrow}
\delta(\rho-\rho_c) \quad ; \quad \rho_c \sim \frac{1}{T_c}
\label{eqn_I} 
\end{equation}

Usual counting tells us that the physics should be a function
of $\theta/N$ and so periodic in $2\pi N$. However $Q$ integer
tells us it is periodic in $2\pi$ for all $N$. This implies
\cite{EWtheta}
that there are in fact $N$ vacua that intertwine as $\theta$
varies. At $\theta =0$ there are $N$ `vacua' which become
stable (exponentially fast) as $N\to\infty$. One can see
these as a continuation from $\mathcal{N}=1$ SUSY of the 
$N$ degenerate vacua associated with gluino condensation
\cite{MStheta}.  
As a first step, on the lattice side, it has been shown
\cite{PisaQ} 
that the moments of $Q$ at $\theta=0$ do suggest a vacuum 
periodicity of $2\pi N$ rather than $2\pi$. What one
would like to see are the near-stable `vacua' directly.
It is easy to calculate their energies
\begin{equation}
E_n(\theta=0)
=
\frac{1}{2}\chi_t (2\pi n)^2 V \quad ; \quad
N\to\infty
\label{eqn_Evac} 
\end{equation}
where $\chi_t \simeq 0.39\surd\sigma \simeq 2T_c/3$.
Thus sometimes one might tumble into an $n{\not=}0$
vacuum from the deconfined phase at $T_c$, but it should
be easier from the bulk phase. This could be done on the
lattice.

\section{Some directions ...}
\label{section_Where}

It would be very nice to have the spectrum of $QCD_{N=\infty}$ 
with light quarks: glueballs and mesons, ground state and
excited, all stable and non-mixing. Since this can be
approached through extrapolating quenched theories in $N$,
this is feasible.

As I pointed out, the intertwining $\theta$-vacua might 
well be directly accessible in lattice calculations.
It would be nice to follow them back to $\mathcal{N}=1$ SUSY
by varying, say, a gluino mass parameter.  

Space-time reduction and the non-analyticities at $N=\infty$
might have important phenomenological and theoretical
implications as conjectured earlier in this talk.

\section*{Acknowledgments}

I am grateful to the organisers for arranging such a productive
meeting and to the ECT, Trento, for the excellent hospitality. Parts
of my write-up have benefitted from my attendance at the ECT
`Hadrons and Strings' workshop that immediately followed the `Large N'
meeting and also from my participation in the `QCD and String Theory'
workshop at the KITP, UCSB, later in the Summer. I am grateful
to the organisers of all these meetings. This write-up has
benefitted from my discussions with many of the participants
at these meetings, in particular
Ofer Aharony, Jose Barbon, David Gross, Rajamanan
Narayanan, Herbert Neuberger, Misha Stephanov and Larry Yaffe.

%
%
%
%

\end{document}